\documentclass[final,3p,11pt]{elsarticle}
\usepackage{geometry}
\usepackage{graphicx}
\usepackage{amsmath}
\usepackage{xcolor}
\usepackage{amsthm}
\usepackage{amssymb}
\usepackage{lineno}
\usepackage{bm}
\usepackage{multirow}
\usepackage{subcaption}
\usepackage{threeparttable}
\usepackage{todonotes}
\usepackage{makecell}
\usepackage{hyperref}
\makeatletter
\def\ps@pprintTitle{%
  \let\@oddhead\@empty
  \let\@evenhead\@empty
  \let\@oddfoot\@empty
  \let\@evenfoot\@oddfoot
}
\makeatother

\begin{document}
\begin{frontmatter}

\title{\textbf{Human-Machine Interaction in Automated Vehicles: Reducing Voluntary Driver Intervention}}
\author[add1]{Xinzhi Zhong}
\author[add2]{Yang Zhou}
\author[add3]{Varshini Kamaraj}
\author[add4]{Zhenhao Zhou}
\author[add1]{Wissam Kontar}
\author[add4]{Dan Negrut}
\author[add3]{John D. Lee}
\author[add1]{Soyoung Ahn\corref{cor1}}
\address[add1]{Department of Civil and Environmental Engineering, University of Wisconsin-Madison}
\address[add2]{Zachry Department of Civil and Environmental Engineering, Texas A$\&$M, College Station}
\address[add3]{Department of Industrial and Systems Engineering, University of Wisconsin-Madison}
\address[add4]{Department of Mechanical Engineering, University of Wisconsin-Madison}
\cortext[cor1]{Corresponding author: sue.ahn@wisc.edu}

\begin{abstract}
This paper develops a novel car-following control method to reduce voluntary driver interventions and improve traffic stability in Automated Vehicles (AVs). Through a combination of experimental and empirical analysis, we show how voluntary driver interventions can instigate substantial traffic disturbances that are amplified along the traffic upstream. Motivated by these findings, we present a framework for driver intervention based on evidence accumulation (EA), which describes the evolution of the driver's distrust in automation, ultimately resulting in intervention. Informed through the EA framework, we propose a deep reinforcement learning (DRL)-based car-following control for AVs that is strategically designed to mitigate unnecessary driver intervention and improve traffic stability. 
Numerical experiments are conducted to demonstrate the effectiveness of the proposed control model.
\emph{The code supporting the findings of this study are available in  \href{https://github.com/XinzhiZhong}{Github page}}. 
\end{abstract}

\end{frontmatter}

\section{Introduction}
The current, and predictably the future, traffic environment is saturated with vehicles equipped with various automation features (between SAE level 2-4 \citep{SAEInternational(2018)}). Most notably, Adaptive Cruise Controller (ACC) has enjoyed large market deployment. Such systems have been extensively studied in the literature, noting their benefits on safety, vehicle emissions, and implications for traffic operations. A critical, yet often overlooked, implication of these technologies is the nuances of human-machine interactions and their impacts on the observed performance and benefits of such technologies. One of those critical interactions is the human control takeovers (often referred to as control disengagement). Yet such disengagement from automation and control takeover by human drivers is ubiquitous \citep{eriksson2017takeover,mcdonald2019toward}. We distinguish here between two kinds of control takeovers: (1) initiated or requested by the automation system, and (2) voluntary takeover by humans even when not critical. In the first, the control takeover is often initiated and requested by the automation system under various scenarios, including safety-critical ones. For such scenarios, a wealth of human factor studies have investigated how efficiently and safely human drivers can assume control of the vehicle. However, case (2) is less studied. Recent empirical evidence from naturalistic driving data found that voluntary takeover by human drivers (when not prompted by the system) is common \citep{DanielNDS,morando2020driver, gershon2021driver}. From \citet{DanielNDS}, it is estimated that a voluntary takeover occurs on average once every four miles. Voluntary takeover is attributed to driver trust in automation, or lack thereof \citep{carneyvoluntary}. Particularly, studies suggest that driver's trust in automation erodes when the automated driving style is dissimilar to the human driving style \citep{ma2021drivers,bellem2018comfort,oliveira2019driving}. 

Regardless of who initiates takeover, control transition could spell trouble for traffic flow stability. Intentional control takeover stems from the necessity or desire to substantially change the course of driving. Thus it can involve a sudden change in driving behavior that could propagate through the traffic stream and cause major traffic disturbances. Notably, it is well-documented that even a subtle deceleration-acceleration movement by a vehicle can eventually develop into a full-blown stop-and-go disturbance in congested, human-driven vehicular traffic \citep{zheng2011freeway,chen2012microscopic}. Traffic disturbances instigated during control transition could be more severe and thus lead to greater traffic instability. 

Clearly, safety comes first, and efficient control transition is sometimes necessary. However, as studies have shown, traffic instability can also have negative safety consequences (e.g., rear-end collisions) \citep{zheng2010impact,makridis2020empirical}. Thus, it is desirable to reduce unnecessary, voluntary driver intervention. A solution to this would be to align the automated driving style with the human driver’s liking. The current ACC systems allow for some customization by enabling the human driver to adjust specific parameters, such as time headway or speed, to their driving preference. However, the alignment between human preference and automated driving style remains largely an afterthought, and takeover implications are yet to be fully studied and designed for.


This paper is then concerned with how voluntary driver intervention impacts traffic and how to mitigate it. Towards this end, the objectives of this study are two-fold: (1) characterize vehicle kinematics during voluntary driver intervention and the ensuing disturbance evolution and (2) develop a deep reinforcement learning (DRL)-based car-following (CF) control framework with multiple objectives, including one to reduce unnecessary driver intervention. As a unique contribution, the DRL-based control is informed by two major factors: (i) evolution of the driver's distrust in automation that leads to eventual driver intervention and (ii) traffic stability. Specifically, (i) is described through evidence accumulation (EA) modeling \citep{kamaraj2022accumulating}, where distrust is influenced by dissimilarity between the automated and human driver CF behavior. (ii) entails the stochastic CF behavior of human drivers, which tends to amplify traffic disturbances. The DRL-based control aims to strike a balance between (i) and (ii) by emulating the human driver's CF behavior while stabilizing traffic.

One may note that reducing driver intervention comes at the expense of driver freedom of action and perhaps has safety implications. We argue that our approach is based on evolution of human-AV interaction, modeled through the EA model. Our approach considers the stochastic deviation of AV driving behavior over time from human driving preference and aims to best align it. Accordingly, our approach only tackles unnecessary interventions to mitigate undesirable traffic disturbances. We later show that, even when unnecessary intervention happens (which is still allowed under our stochastic framework), the outcome is still better than without our control. Accordingly, we emphasize studying interventions from the lens of traffic-level implications, as it promotes safety and human-AV alignment.

The rest of the paper is organized as follows. In Section \ref{section_ea}, we characterize empirical voluntary driver invention and analyze its impact on vehicle kinematics and disturbance evolution. In Section \ref{section_av}, we introduce a novel DRL-based CF control framework to minimize the driver intervention and its subsequent impact on traffic flow. In Section \ref{performance}, we analyze the performance of our DRL-based control model. Section \ref{conclusion} provides some discussions and concluding remarks.

\section{Driver Intervention and Its Impact on Traffic}\label{section_ea}
This section characterizes the driver intervention behavior through EA modeling and the ensuing disturbance, in terms of vehicle kinematics, using empirical data from a driving simulator experiment. We further examine the evolution of the disturbance through a platoon of human-driven vehicles (HDVs) via numerical simulations.

\subsection{Evidence Accumulation Modeling for Driver Intervention}
Here, we introduce the EA model by \citet{kamaraj2022accumulating} adopted in this study. It postulates that driver $i$ accumulates "evidence", $E_i(t)$, over time $t$ according to the difference between the automated CF control and their manual CF behavior. The driver decides to take over control when $E_i(t)$ reaches a certain threshold, $E_i^{T}$. This process is modeled with a drift-diffusion modeling framework as shown in Eqs. (\ref{EA})-(\ref{indicator}). Note that we only model the first driver-initiated takeover and assume that the driver does not transition back to automated mode after the switch to manual driving mode.

\begin{flalign}\label{EA}
    &E_i(t)=E_{i}(t-1) + d_{i}U_{i}(t)\eta_i(t) + \alpha \epsilon(t)
\end{flalign}
\begin{flalign}\label{Dissimilarity}
   U_{i}(t) &= {w_1}_i\Phi_{i,x}(t)+{w_2}_i\Phi_{i,v}(t)+{w_3}_i \Phi_{i,TTA}(t)
\end{flalign}

\begin{flalign}\label{Normalized_X}
   \Phi_{i,x}(t)&= \frac{\sqrt{(\Delta x_{i,AV}(t)-\Delta x_{i,HDV}(t))^2} - \varphi^{min}_x}{\varphi^{max}_x - \varphi^{min}_x}
\end{flalign}
\begin{flalign}\label{Normalized_V}
   \Phi_{i,v}(t) &= \frac{\sqrt{(\Delta v_{i,AV}(t)-\Delta v_{i,HDV}(t))^2} - \varphi^{min}_v}{\varphi^{max}_v - \varphi^{min}_v}
\end{flalign}
\begin{flalign}\label{Normalized_A}
   \Phi_{i,TTA}(t) &= \frac{max(0, {TTA_i}_R(t) - {TTA}_A(t)) - \varphi^{min}_{TTA}}{\varphi^{max}_{TTA} - \varphi^{min}_{TTA}}
\end{flalign}

\begin{flalign}\label{indicator}
    &\eta_i(t)=
  \begin{cases}
      1 & E_i(t) \leq E_i^{T} \\
      0 & E_i(t) > E_i^{T} \\
   \end{cases}
\end{flalign}
where $d_i$ is the drift rate representing the rate at which driver $i$ accumulates the evidence. $E(0)$ is the starting point of the accumulated evidence. This parameter can reflect if drivers are biased in their trust toward the automation. For example, distrusting drivers may have higher starting values which could prompt quicker driver-initiated transitions to manual control. $U_{i}(t)$ represents the difference in driving style, simplified as the weighted sum of (1) the normalized root square error (NRSE) between the relative spacing maintained by the AV, $\Delta x_{i,AV}(t)$, and the relative spacing that would be maintained by the human driver, $\Delta x_{i,HDV}(t)$; (2) NRSE between the relative speed $\Delta v_{i,AV}(t)$ and $\Delta v_{i,HDV}(t)$; (3) NRSE between the estimated travel time (i.e., maintaining the current speed), ${TTA_i}_R$ and the target travel time, ${TTA}_A$. (1) and (2) provide evidence of dissimiliarity between the automation's CF behavior and drivers' manual CF behavior. (3) provides evidence regarding how driving time pressure influences the level of distrust. We apply the min-max normalization to these three terms, $\varphi^{max}$ and $\varphi^{min}$ are the minimum and maximum scale factors. ${w_1}_i$, ${w_2}_i$ and ${w_3}_i$ are the weights that $\sum_{j=1}^3{w_j}_i=1$. $\epsilon(t)$ is the white Gaussian noise with variance $\sigma^2$ added to the evidence for diffusion. $\alpha$ is an adjustment coefficient for the diffusion process. The $\alpha$ is recommended as $0.1$ \citep{palada2016evidence,abut2023towards}. $\eta_i(t)$ is the binary variable for the driving mode, $1$ for automated mode and $0$ for human driving mode. As noted, $\Delta x_{i,AV}(t)$, $\Delta v_{i,AV}(t)$ for the automated mode is determined by the AV control, $f_{i,AV}$, and $\Delta x_{i,HDV}(t)$, $\Delta v_{i,HDV}(t)$ by the human driver, $f_{i,HDV}$.

Then we generalize the CF law, with human intervention, by integrating the EA model.

\begin{flalign}\label{CF_law}
    \ddot{y}_i(t)&=(1-\eta_i(t))f_{i,HDV}(t)+\eta_i(t)f_{i,AV}(t)
\end{flalign}
where $f_{i,HDV}(t)=f_{i,HDV}(y_{i-1}(t),\dot{y}_{i-1}(t),y_{i}(t),\dot{y}_{i}(t),\theta_{i,HDV})$ represents the CF behaviors of HDVs, and $f_{i,AV}(t)=f_{i,AV}(y_{i-1}(t),\dot{y}_{i-1}(t),y_{i}(t),\dot{y}_{i}(t),\theta_{i,AV})$ represents the CF control of AVs. ${y}_{i}(t)$, $\dot{y} _{i}(t)$, and $\ddot{y}_{i}(t)$ respectively represent the kinematics characteristics (i.e., the position, speed, and acceleration) at time $t$ for vehicle $i$. $\theta_{i,HDV}$ and $\theta_{i,AV}$  are CF model parameters, for HDV and AV, respectively. By Eq. (\ref{Dissimilarity}), we can further represent $\Delta v_{i,AV}$ and $\Delta v_{i,HDV}$ by Eqs. (\ref{relative_spacing_av})-(\ref{relative_tta_av}):

\begin{flalign}\label{relative_spacing_av}
    &\Delta x_{i,AV}(t)={y}_{i-1}(t)-{y}_{i,AV}(t)
\end{flalign}
\begin{flalign}\label{relative_spacing_hdv}
    &\Delta x_{i,HDV}(t)={y}_{i-1}(t)-{y}_{i,HDV}(t)
\end{flalign}
\begin{flalign}\label{relative_speed_av}
    &\Delta v_{i,AV}(t)={\dot{y}}_{i-1}(t)-{\dot{y}}_{i,AV}(t)
\end{flalign}
\begin{flalign}\label{relative_speed_hdv}
    &\Delta v_{i,HDV}(t)={\dot{y}}_{i-1}(t)-{\dot{y}}_{i,HDV}(t)
\end{flalign}
\begin{flalign}\label{relative_tta_av}
    &{TTA_i}_R(t)=\frac{D_T-{y}_{i,AV}(t)}{{\dot y}_{i,AV}(t)}
\end{flalign}
where $D_T$ is the total travel distance.

The generalized CF law (Eq. (\ref{CF_law})) will be integrated in our DRL-based CF control later on. 

\subsection{Model Calibration through Driving Simulator Experiment}
\begin{figure*}[h]
        \captionsetup{justification=centering}
 \includegraphics[width=\textwidth]{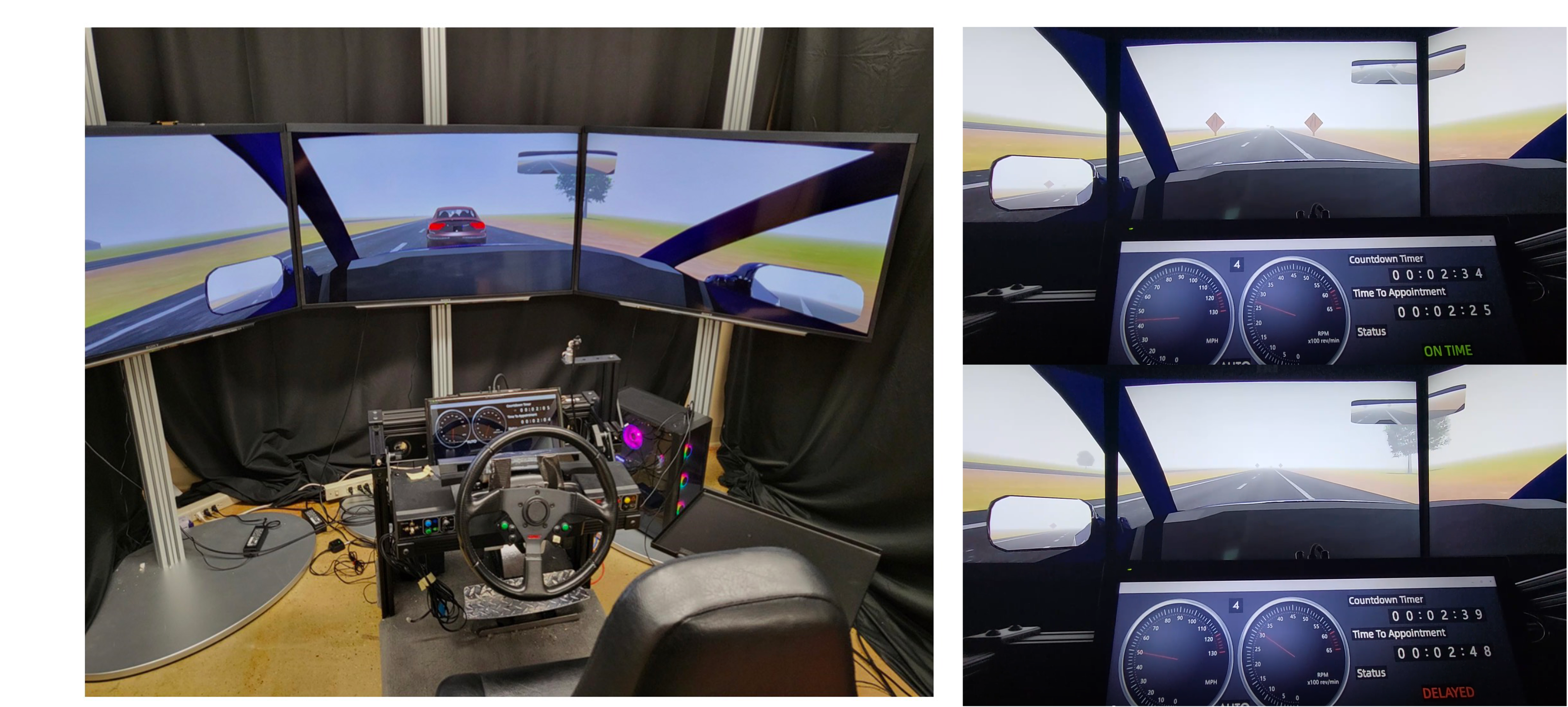}
    \caption[Driving Simulator Experiment]%
        {\small Driving Simulator Experiment\\}
\label{simulator}
\end{figure*}

In this subsection, we estimate the driver intervention behavior by calibrating the EA model in Eqs. (\ref{EA})-(\ref{indicator}) using data from an experiment conducted in a driving simulator. Specifically, a fixed-based driving simulator (Fig. \ref{simulator}) was used to collect the driving data associated with control transitions between automated and manual driving. The configuration of automated driving is grounded in the intelligent driver model (IDM) and supplemented by a proportional-integral-derivative (PID) controller to govern the accelerator and brake inputs within Chrono::HIL \citep{zhou2024open,zhou2023chrono}. Chrono::HIL is a traffic simulation middleware with human-in-the-loop support built upon Project Chrono, an open-source multi-physics simulation engine \citep{tasora2016chrono} which provides simulation support of vehicle dynamics \citep{serban2023real}, visual rendering, hardware interfacing and simulation data write-out functionalities. The simulator is located at the University of Wisconsin-Madison. The fixed-based driving simulator includes a three-screen setup for driver-view rendering, a dashboard monitor for tachometer and speedometer simulations, and a timing section displaying trip details like time-to-destination and duration. The timing section helps drivers determine if they are scheduled to arrive on time or if there may be delays. It also has a fixed-base chassis equipped with a force feedback steering wheel and pedals.

\begin{figure}[h]
    \centering
        \includegraphics[width=0.9\linewidth]{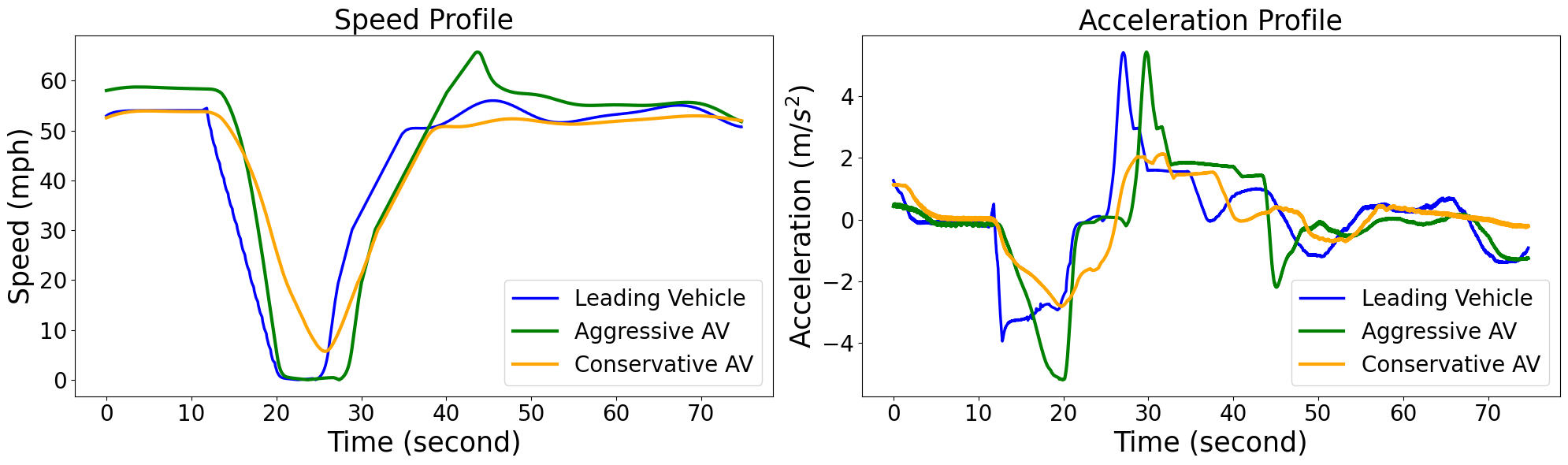}
    \caption{Experimental Setting in the Driving Simulator}
    \label{fig_automated}
\end{figure}

A total of 48 participants were recruited. In the experiment, drivers traveled on a two-lane highway and encountered road construction in the left lane, and were presented with a leading vehicle before entering the construction zone. The speed and acceleration profiles of the leading vehicle are shown in Fig. \ref{fig_automated}. Each participant drove in the simulator three times. They first drove the simulator vehicle in purely manual mode to establish their manual driving behavior. Then, they experienced two types of automated CF driving styles – aggressive and conservative; see Fig. \ref{fig_automated} for their speed and acceleration profiles. In these modes, drivers were permitted to switch between manual and automated modes at will, by pressing a button on the steering wheel that either activated automated mode or deactivated automated mode. Each drive lasted approximately 3 minutes.  

During the experiment, drivers were asked to consider each drive as a trip to a work-related meeting and asked to arrive in under 3 minutes while obeying traffic laws. Traffic laws were defined using the suggested speed limit ($60$ mph) and safe driving behavior. Notably, engaging in a non-driving task while in manual mode constituted unsafe driving behavior. Successful completion of the drive meant arriving safely in under 3 minutes while always driving at or below the suggested speed limit if driving in manual mode. If the drive was completed successfully, drivers receive a cash bonus. If the drive was unsuccessful, the bonus was withheld either partially in the case of late arrivals or fully in the case of speeding and crash events. This combination of experimental conditions ensured that drivers could (1) engage in manual CF, (2) observe automated CF, and (3) initiate takeover depending on their preference.

The investigation starts from the equilibrium states, where the speeds of the leader and the follower remain unchanged before the deceleration of the leader, excluding the initial acceleration phrase from $0$ to the equilibrium speed. We observed a total of 34 takeover events out of the 48 scenarios involving the automation modes, resulting in a takeover probability of approximately 70.83\%. Of those, 19 and 15 takeover instances were initiated under the conservative and aggressive automation settings, respectively, and 14 instances were initiated with no takeovers.  5 instances were further excluded. 4 of them include lane changes occurring in manual driving mode, as this complicates the driving dissimilarity measure. 1 instance includes a driver takeover right at the start (i.e., at the initial acceleration phrase). Thus we have 29 interventions out of 44 clean instances for further analysis under the CF scenarios.

$E_i(0)$, $d_i$, $E_i^{T}$, ${w_1}_i$, ${w_2}_i$ and ${w_3}_i$ represent the individual differences in driver behavior such as prior bias towards automation use. Thus, we calibrate these parameters in a stochastic fashion to accommodate the driver heterogeneity using the Approximate Bayesian Computation (ABC) with an adaptive sequential Monte Carlo sampler (ABC-ASMC) \citep{del2012adaptive,zhong2023understanding}. This method is simulation-based and allows one to estimate the empirical posterior joint distribution of parameters in a likelihood-free fashion. Specifically, the main principle of the ABC method is as follows: 1) sample "particles" (a set of EA model parameter values) from assumed prior distributions; 2) simulate the evolution of distrust per Eq. (\ref{EA}) based on the particles; 3) evaluate the closeness between the observed and simulated takeover points based on a goodness of fit measure (GOF); 4) accept particles within the tolerance; and 5) approximate the posterior distributions using the accepted particles. 

\begin{figure}[!htb]
    \centering
    \captionsetup{justification=centering}
    \begin{subfigure}{0.48\textwidth}
        \centering
        \includegraphics[width=\linewidth]{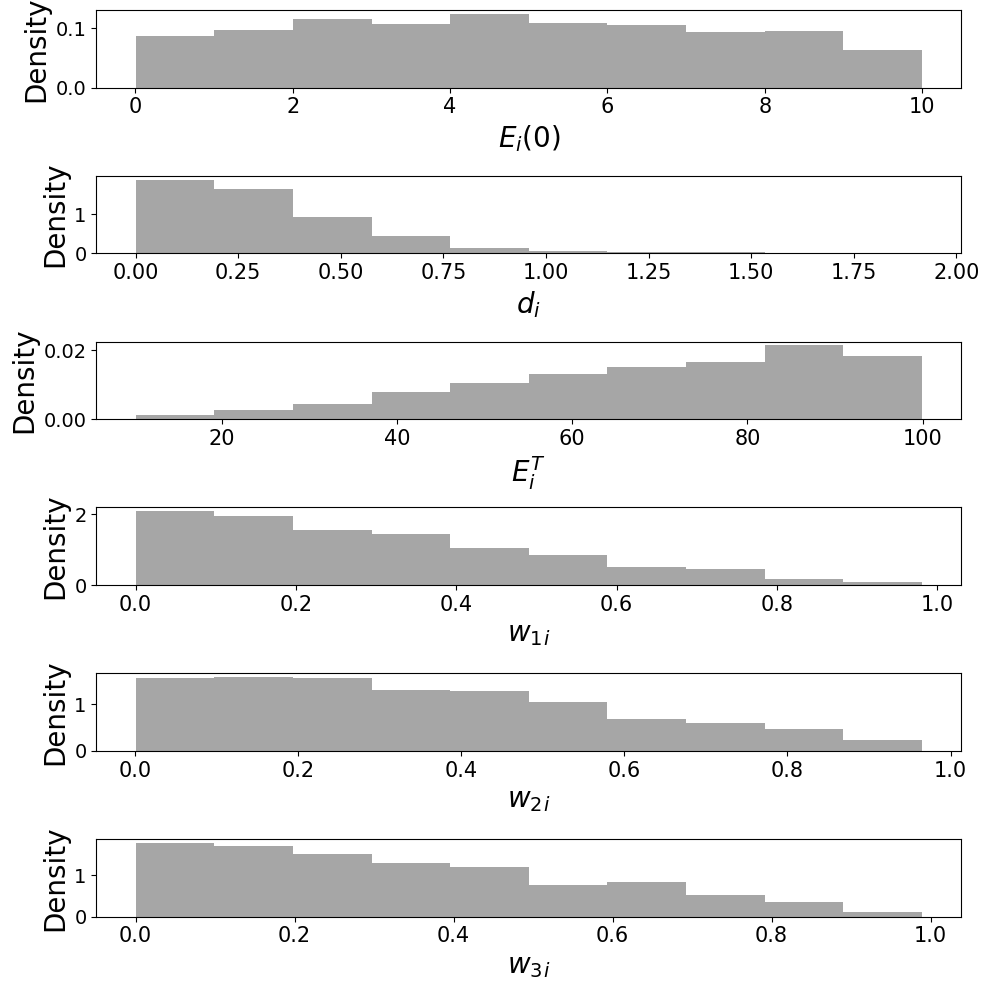}
        \caption{}
        \label{fig:figure1}
    \end{subfigure}
    \hfill
    \begin{subfigure}{0.48\textwidth}
        \centering
        \includegraphics[width=\linewidth]{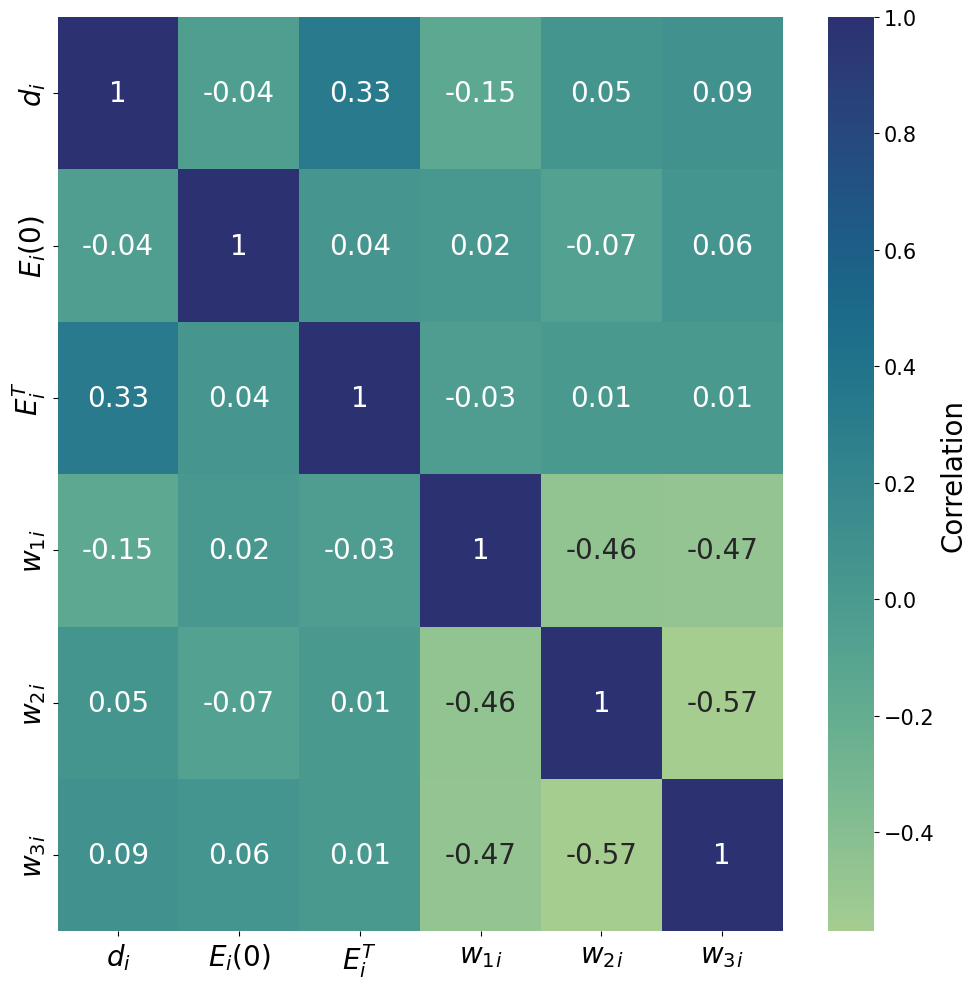}
        \caption{}
        \label{fig:figure3}
    \end{subfigure}   
    \caption{EA models calibrated by ABC-ASMC\\
    (a) Marginal Distribution (b) Correlation Matrix }
    \label{EA_Calibration}
\end{figure}

Note that the driving simulator collects trajectory data at the resolution of 0.01 seconds --- too high for a human driver to perceive any difference. Thus, we set the time step $t$ for the EA model to a more perceptible 0.1 seconds. Due to the lack of prior knowledge, the prior distributions for the model parameters are set as independent uniform distributions, whose marginal distributions are set as: $d\sim Uniform(0,2)$\citep{wagenmakers2007ez}, $E(0)\sim Uniform(0,10)$, $E^{T}\sim Uniform(10,100)$, $w_{i,1} \sim Uniform(0,1)$, $w_{i,2} \sim Uniform(0,1)$. And the estimated posterior marginal distributions and correlation among the parameters are shown in Fig. \ref{EA_Calibration}. The distribution of $E(0)$ is largely uniform across its range. Further, it shows a weak correlation with $d_i$ and $E_i^T$, implying that $E(0)$ has little impact on the driver's propensity to intervene. In contrast, $d_i$ and $E_i^{T}$ show stronger positive correlation. This suggests that as the drift rate becomes larger, the evidence threshold leading to intervention also increases. $d_i$ is negatively correlated with ${w_1}_i$, weakly correlated with ${w_2}_i$, but positively correlated with ${w_3}_i$. It indicates the accumulation process tends to be slow when more evidence is collected based on the relative spacing difference, whereas the task pressure appears to accelerate the process. The weights exhibit a strong negative correlation, suggesting that the evidence is predominantly collected by one of them.

\begin{figure*}[!htb]
        \centering
        \captionsetup{justification=centering}
        \begin{subfigure}[b]{0.45\textwidth}
 \includegraphics[width=\textwidth]{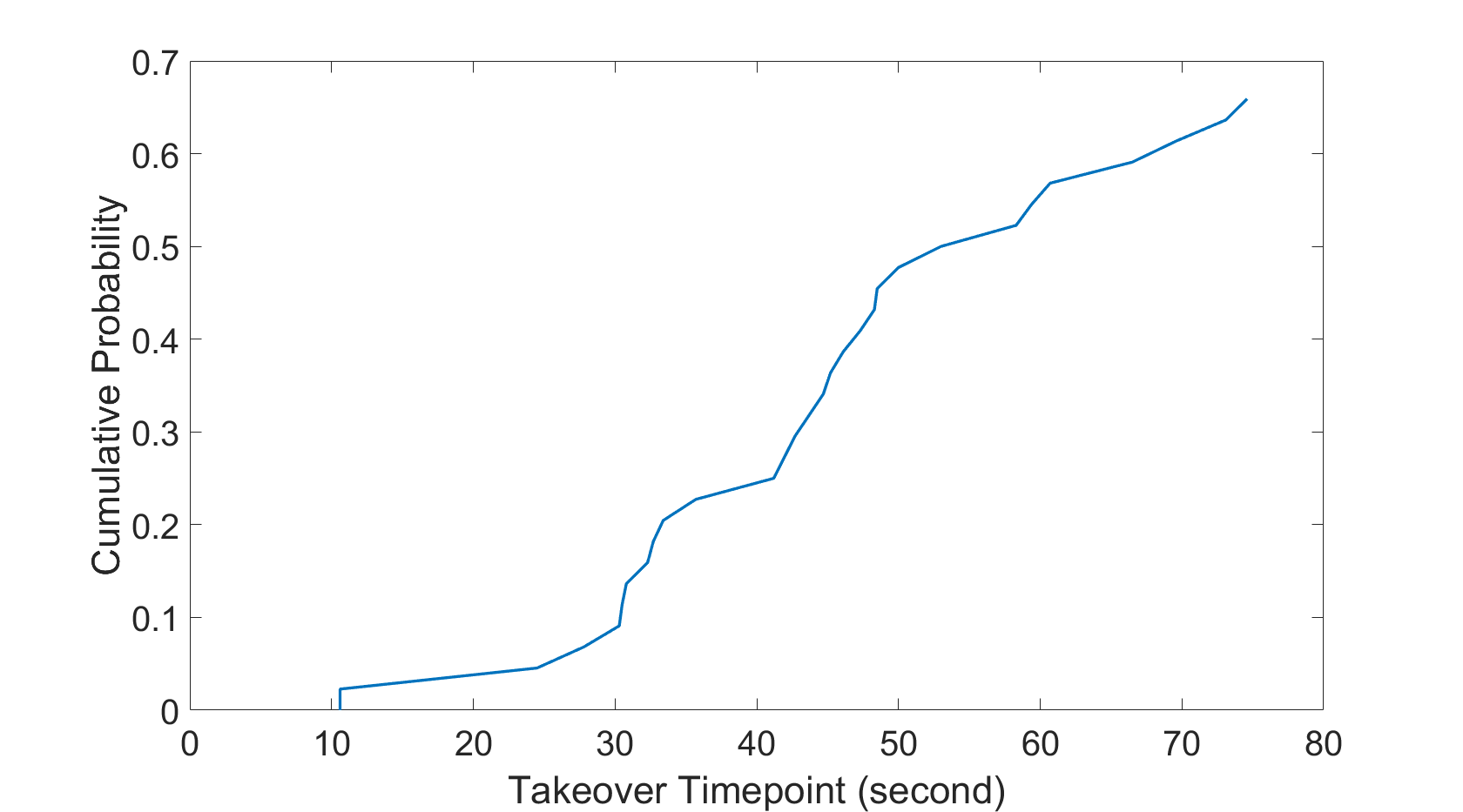}
              \caption[]%
            {{\small }}   
        \end{subfigure}
        \hspace{1em}
        \begin{subfigure}[b]{0.45\textwidth} 
            \centering \includegraphics[width=\textwidth]{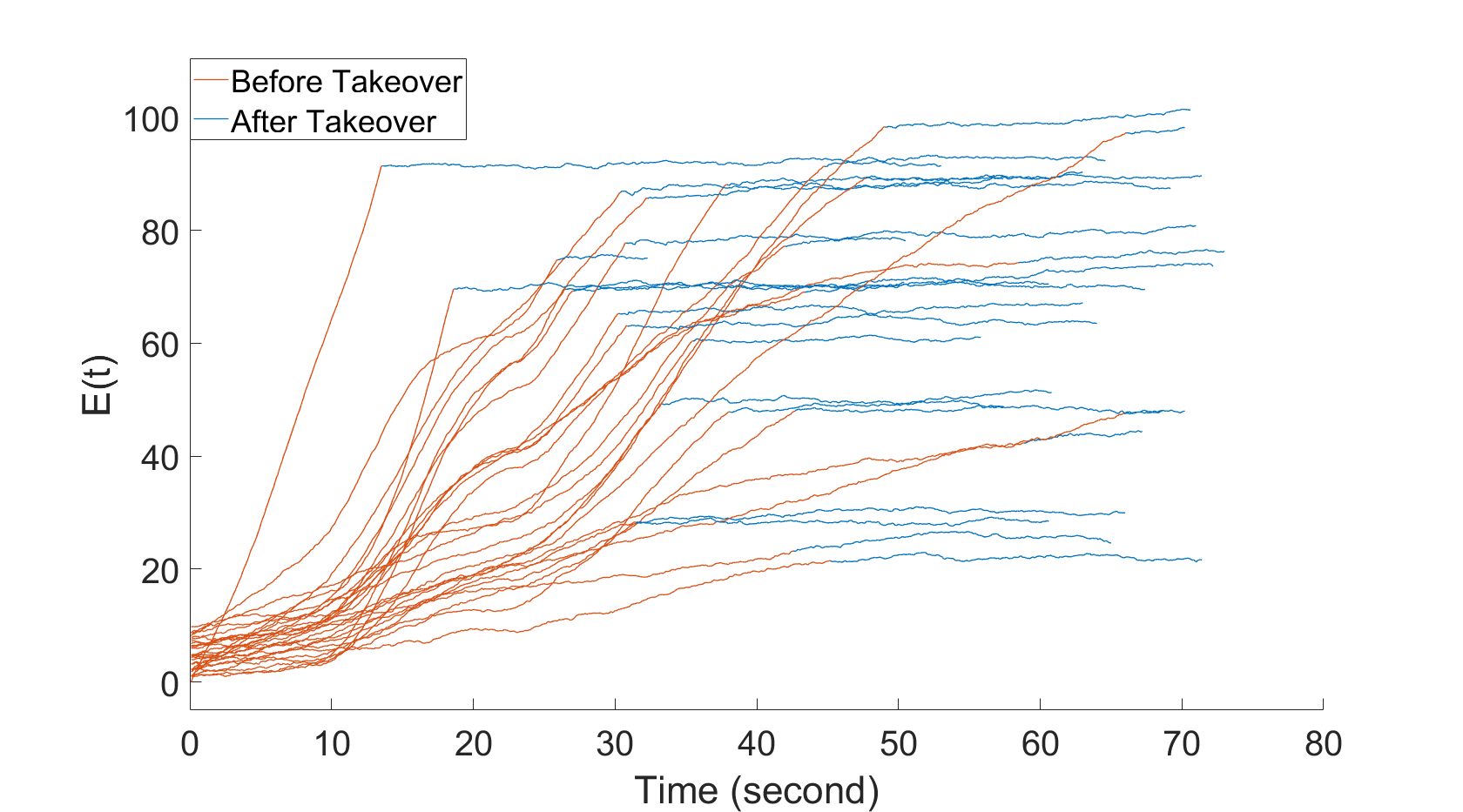}
                 \caption[]%
            {{\small }}   
        \end{subfigure}
         \caption[Evidence Accumulation of Driving Simulator Experiments]%
        {\small Evidence Accumulation in the Driving Simulator\\
       (a) Cumulative Distribution of Takeover Time-point (b) Evidence Accumulation for Intervention Events}
        \label{emp_takeover}
    \end{figure*}

Fig. \ref{emp_takeover}(a) and \ref{emp_takeover}(b) respectively show the estimated evidence accumulation processes for the instances based on the best-fitted particles (Fig. \ref{EA_Calibration}) and the empirical cumulative distribution of control takeover time. A surge in control takeover is evident between 30 and 50 seconds, when vehicles are recovering from the disturbance designed in the experiment.

\subsection{Effects of Driver Interventions on Vehicle Kinematics and Disturbance Evolution}
Fig. \ref{impact_vk} shows typical examples of the changes in vehicle kinematics (red rectangle) due to driver intervention from the driving simulator experiments. Specifically, the top-left subplots (i.e., evidence accumulation) show the evolution of $U_{i}(t)$: the dissimilarity between driving behaviors of AV and HDV grows over time. The other three display the deviation from purely automated driving (orange) in terms of position, speed, and acceleration, following the transition to manual driving mode (green). From Fig. \ref{impact_vk}(a-c), we observe a sharp deceleration and speed drop when the driver assumes control of the vehicle. This change instigates a substantial traffic disturbance, visible within the vehicle's trajectory. Once intervening, the trajectory of human driver (green curve) lags behind the trajectory if the intervention did not occur (orange dashed curve). Such an occurrence is inevitable, regardless of whether the autonomous vehicle is set to aggressive or conservative mode. Changes become more pronounced when the takeover happens during the vehicle's acceleration phase, as it recovers from oscillation (Fig. \ref{impact_vk}(c)).

\begin{figure*}[!htb]
        \centering
        \captionsetup{justification=centering}
        \begin{subfigure}[b]{\textwidth}
          \centering 
 \includegraphics[width=0.9\textwidth]{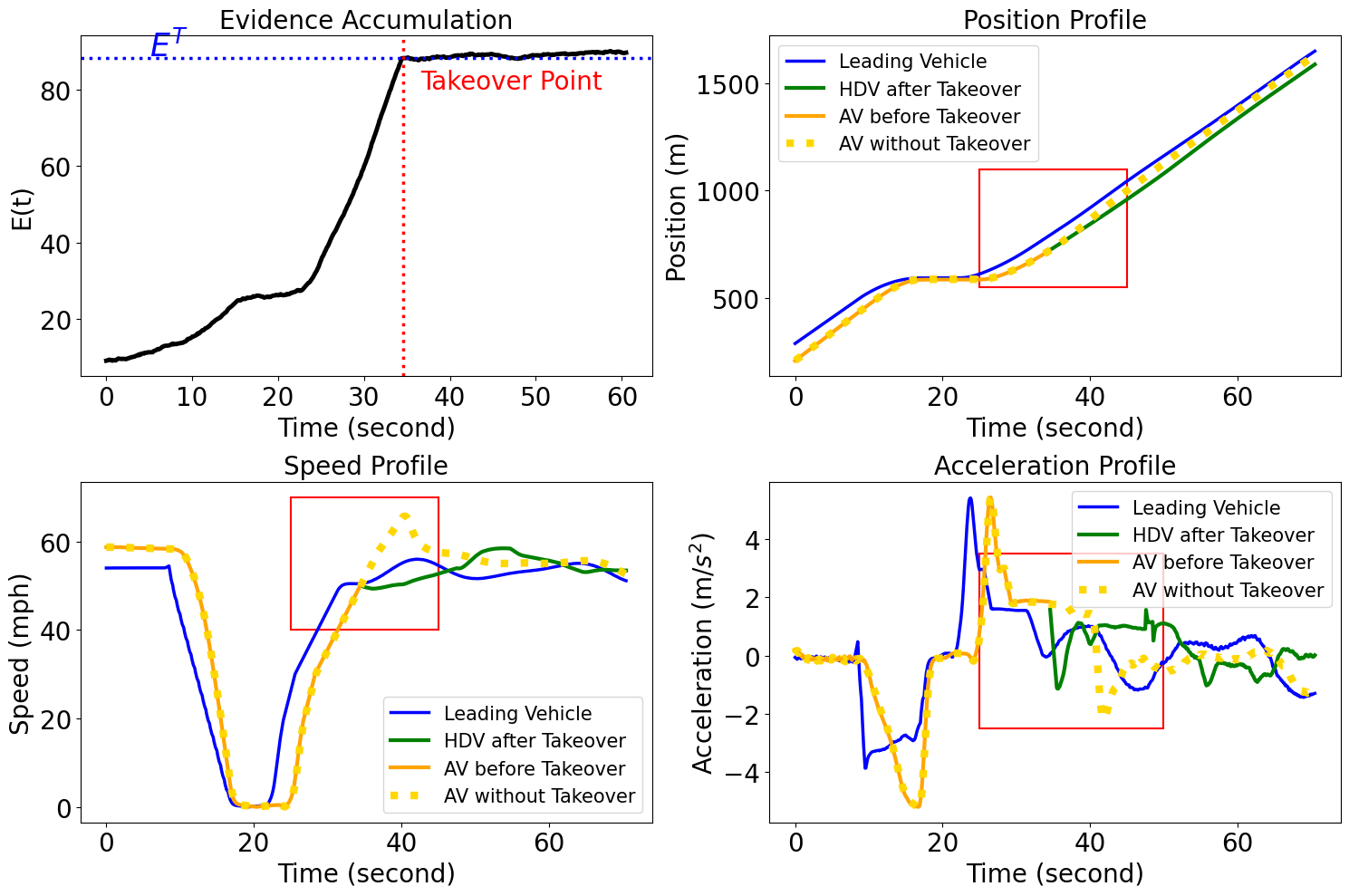}
  \label{impact_vk:(a)}
              \caption[]%
            {{\small }}   
        \end{subfigure}
        \begin{subfigure}[b]{\textwidth} 
            \centering \includegraphics[width=0.9\textwidth]{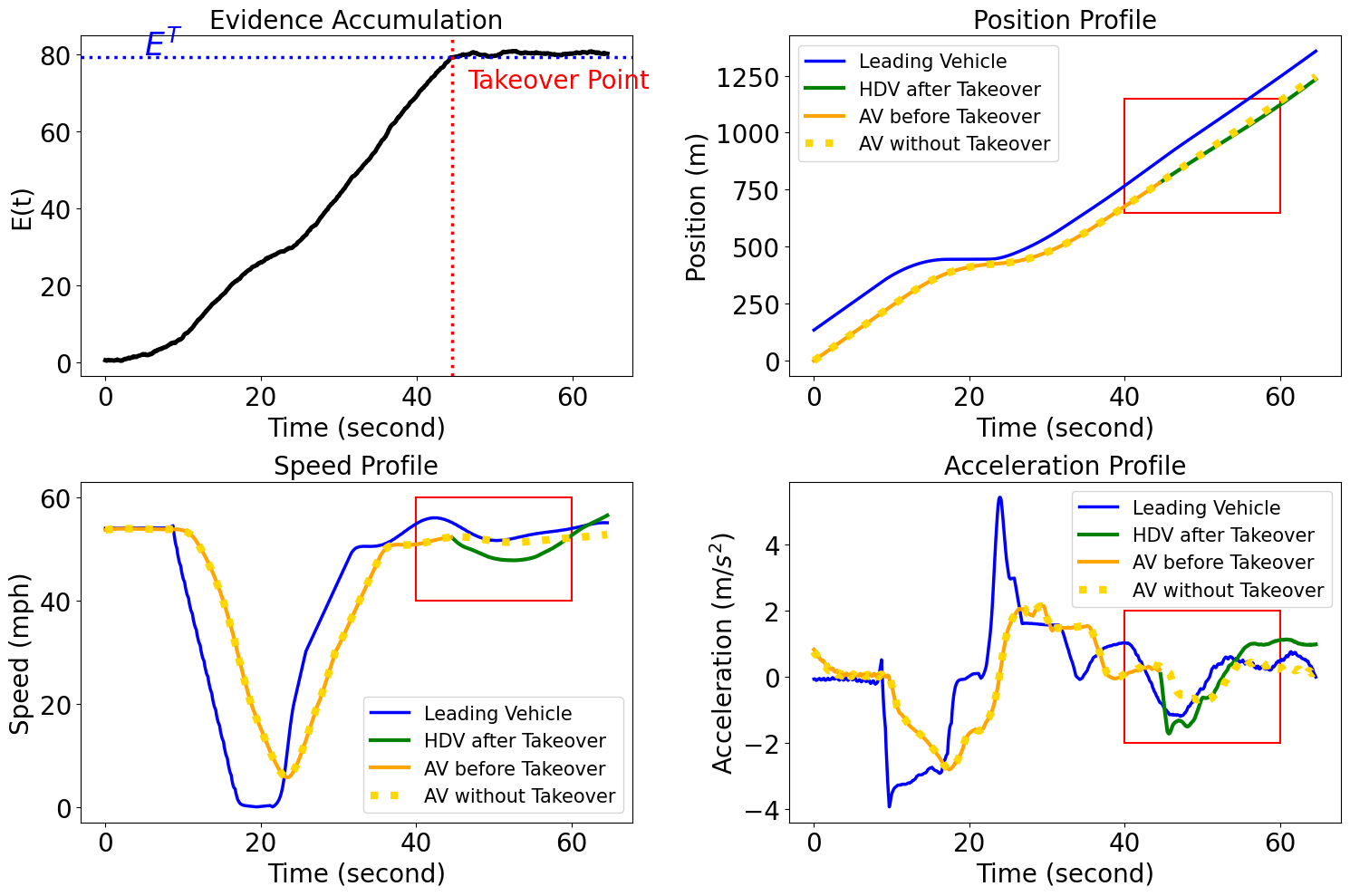}
             \label{impact_vk:(b)}
                 \caption[]%
            {{\small }}   
        \end{subfigure}
\end{figure*}
\begin{figure*}[!htb]
\captionsetup{justification=centering}
    \ContinuedFloat
    \centering
    \begin{subfigure}[b]{\textwidth} 
            \centering \includegraphics[width=0.9\textwidth]{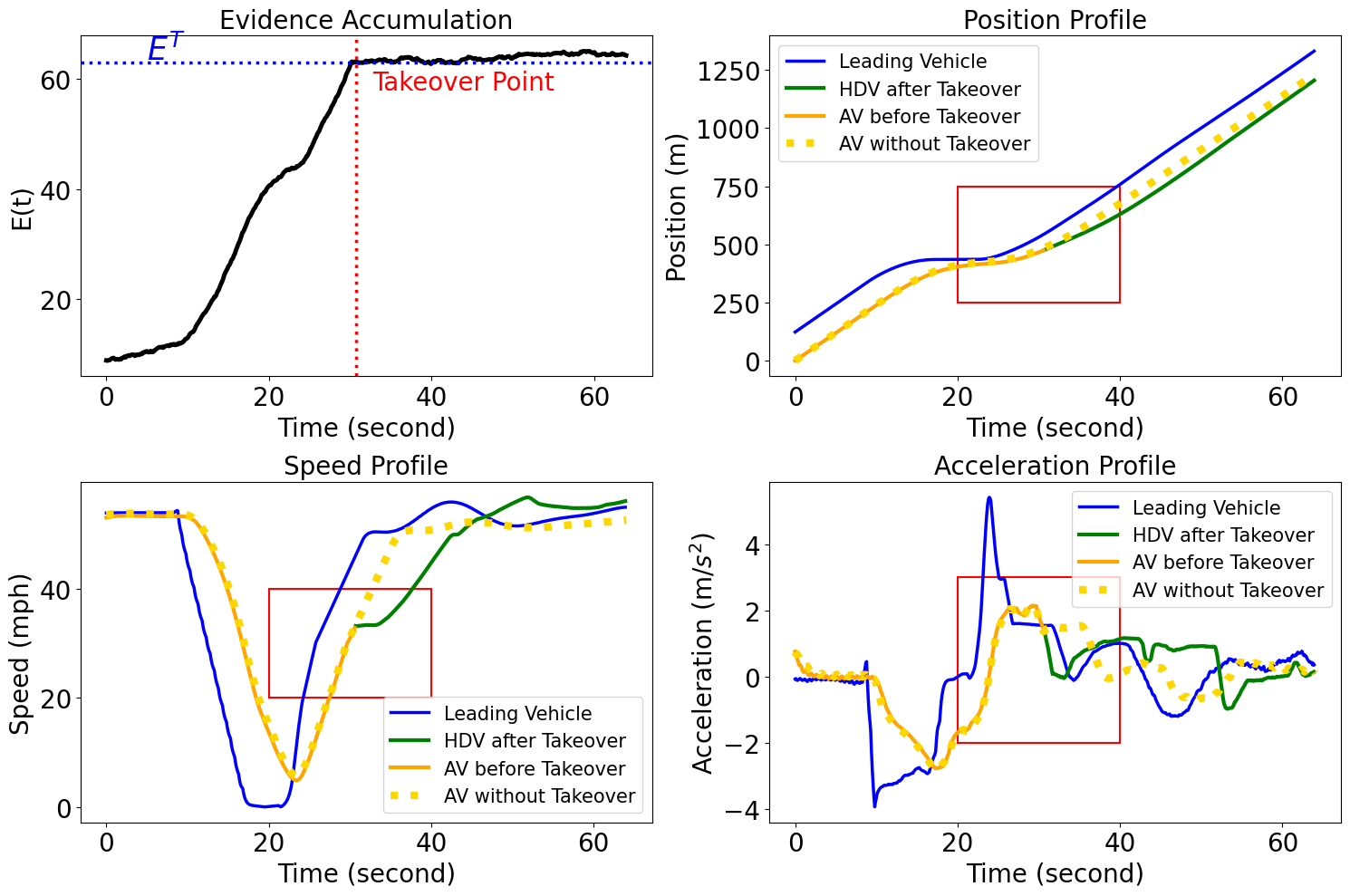}
                 \label{impact_vk:(c)}
                 \caption[]%
            {{\small }}   
        \end{subfigure}
    \caption[Impact of Driver Intervention on Vehicle Kinematics]%
        {\small Impact of Driver Intervention on Vehicle Kinematics from Driving Simulator\\
       (a) Typical Example in Aggressive Automation Setting  \\(b) Typical Example in Conservative Automation setting\\
(c) Example of Driver Takeover During Acceleration Phase}
\label{impact_vk}
\end{figure*}

Having empirically observed that a takeover initiates a disturbance, we extended our investigation by simulating five following HDVs to examine how such a disturbance propagates through the platoon. Specifically, we assume that the followers are HDVs obeying the stochastic hybrid CF model \citep{jiang2023generic}. We use this model as it is designed to accommodate the stochasticity in the real-word HDV behaviors and demonstrates a robust capacity to reproduce the behaviors. In brief summary, this model probabilistically concatenates a pool of CF models based on their relative likelihood of describing observed behavior. The relative likelihood is determined in a data-driven fashion via ABC by comparing the share of accepted particles across the pool of models. The posterior distribution of the hybrid model is then updated based on the relative likelihood. The readers are referred to \cite{jiang2023generic} for more details. As per their calibration results using the NGSIM data \citep{NGSIM}, we assign the probability share of 0.1/0.1/0.7/0.1 respectively among the four well-known CF models: intelligent driver model (IDM) \citep{kesting2010enhanced}, full velocity difference model (FVDM) \citep{jiang2001full}, generalized force model (GFM) \citep{helbing1998generalized} and optimal velocity model (OVM) \citep{bando1995dynamical}. We applied this model to simulate each follower by randomly generating a CF model (among the above four) and a set of model parameter values from the posterior distribution \citep{jiang2023generic}. We postulate the equilibrium speed $\dot{y}_E$ of the first follower sets the standard for the simulated platoon (i.e., 26.2 m/s). The initial positions of the subsequent followers are determined by their standstill spacing to maintain this equilibrium speed, thereby ensuring they all start from the equilibrium state.

\begin{figure*}[!htb]
        \centering
        \captionsetup{justification=centering}
         \begin{subfigure}[b]{\textwidth}
        \centering
        \includegraphics[width=0.475\linewidth]{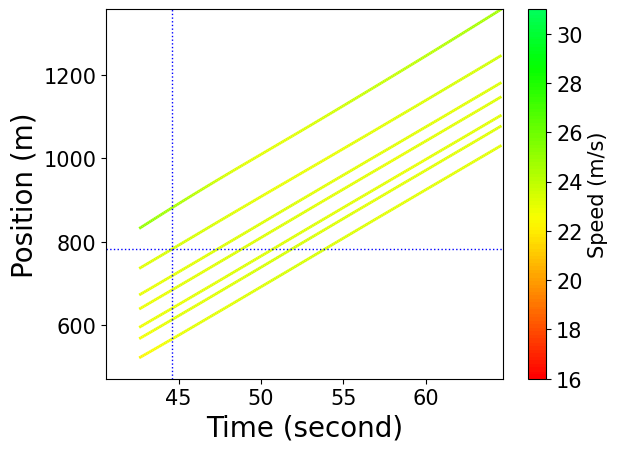}%
        \hfill
        \includegraphics[width=0.475\linewidth]{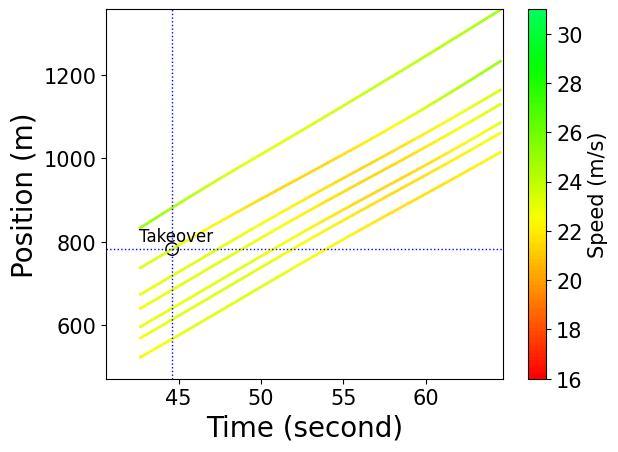}
        \caption{}
    \end{subfigure}
        \begin{subfigure}[b]{\textwidth}
        \centering
        \includegraphics[width=0.475\linewidth]{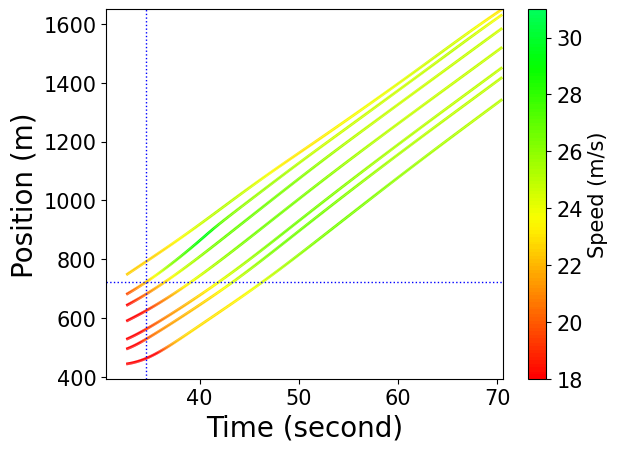}%
        \hfill
        \includegraphics[width=0.475\linewidth]{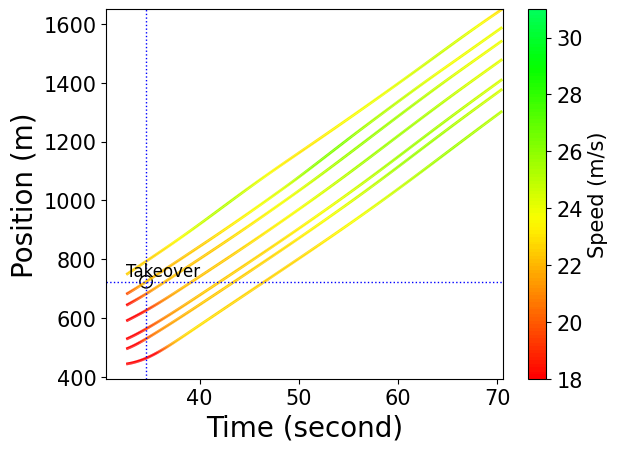}
        \caption{}
    \end{subfigure}
         \hspace{1em}
        \begin{subfigure}[b]{\textwidth}
        \centering
        \includegraphics[width=0.475\linewidth]{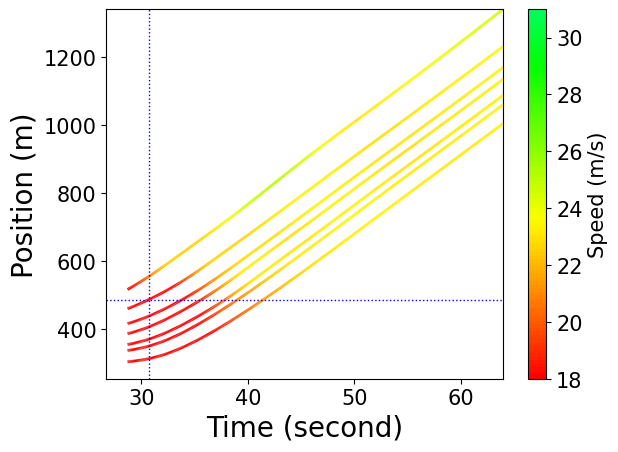}%
        \hfill
        \includegraphics[width=0.475\linewidth]{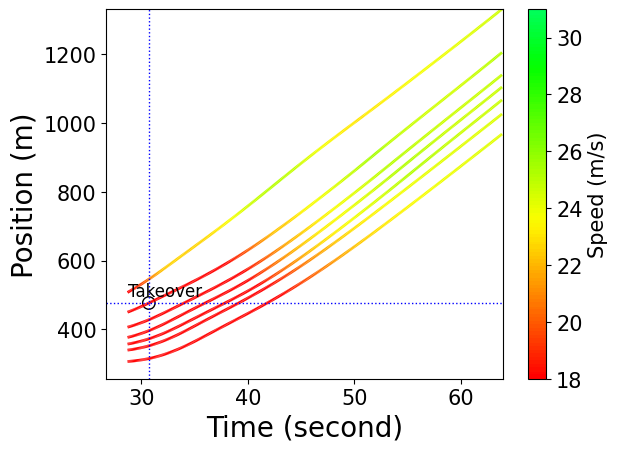}
        \caption{}
    \end{subfigure}
         \caption[Impact of Disturbance Evolution and Propagation from Driving Simulator]%
        {\small Impact on Disturbance Evolution and Propagation \\
       (a) Typical Example in Aggressive Automation Setting  \\(b) Typical Example in Conservative Automation setting\\
(c) Example of Driver Takeover During Vehicle Recovery from Disturbance}
        \label{emp_disturbance}
    \end{figure*}

\begin{figure*}[!htb]
        \centering
\captionsetup{justification=centering}
        \begin{subfigure}[b]{0.45\textwidth}
 \includegraphics[width=0.9\textwidth]{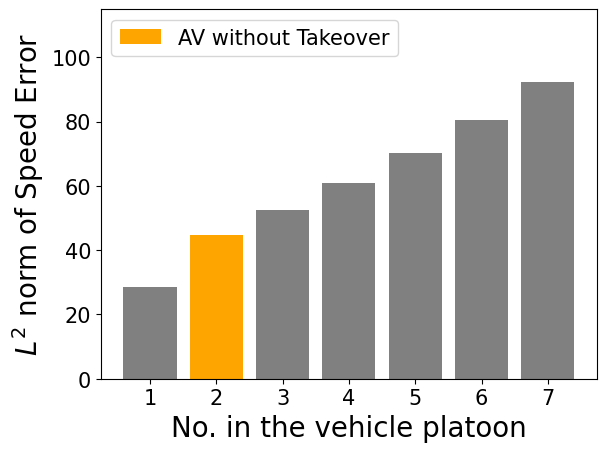}
  \caption*{(a1)}%
            {{\small }}   
        \end{subfigure}
        \begin{subfigure}[b]{0.45\textwidth} 
            \centering \includegraphics[width=0.9\textwidth]{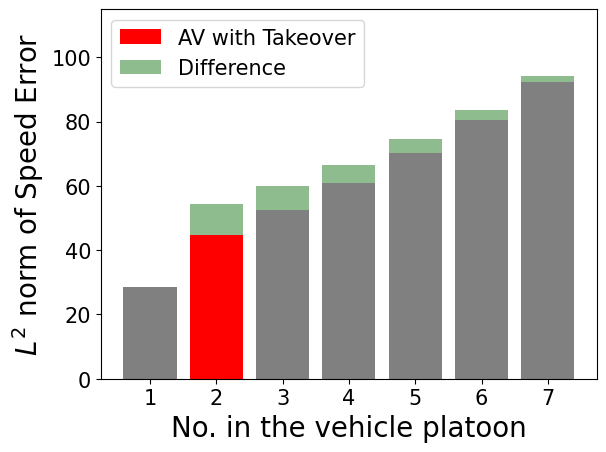}
             \caption*{(b1)}%
            {{\small}}   
        \end{subfigure}
         \caption[ $L_2$ norm of Speed Error]%
        {\small Expectation of $L_2$ Norm of Speed Error over 3000 Simulations\\
        (a1) Without Takeover (b1) With Takeover }
        \label{emp_stability}
    \end{figure*}

Fig. \ref{emp_disturbance} depicts disturbance evolution corresponding to the examples provided in Fig. \ref{impact_vk}, with a particular emphasis on the changes that occur after the takeover point. The left column of Fig. \ref{emp_disturbance} shows the evolution without intervention and the right column with intervention. It is evident from Fig. \ref{emp_disturbance}(a-c) that the evolution of disturbance due to the intervention is pronounced in each case. 
Based on $L_2$ norm string stability \citep{naus2010string,treiber2013traffic}, we further characterize the stability through the platoon by computing the squared $L_2$ norm of the speed error between the equilibrium speed and the speed for each vehicle upstream $\|\dot{y}_i(t)-\dot{y}_E\|_2$, to measure the magnitude of disturbance. Drawing upon the 29 interventions, we replicate the simulation of the five followers 3000 times both with and without intervention. Each value shown in Fig. \ref{emp_stability} is an average over the total simulation runs. The result shows that with driver intervention, the $L_2$ norm for the AV increases significantly, suggesting the instigation of disturbances, and the disturbances grow much more quickly through upstream HDVs.  The result underscores the importance of mitigating voluntary driver intervention when it is not safety-critical.

\section{AV Longitudinal Control to Reduce Impacts of the Interventions}\label{section_av}

Based on the insights from Section \ref{section_ea}, we design an AV control strategy that minimizes unnecessary voluntary driver intervention. Our control design is based on a DRL framework that aims to balance (1) accumulation of distrust in automation by drivers and (2) disturbance amplification. The proposed DRL approach is flexible, allowing us to handle multiple competing objectives and the complexities arising from stochastic driver behavior and traffic dynamics. 

\begin{figure}[!htb]
	\centering
\includegraphics[width=0.6\textwidth]{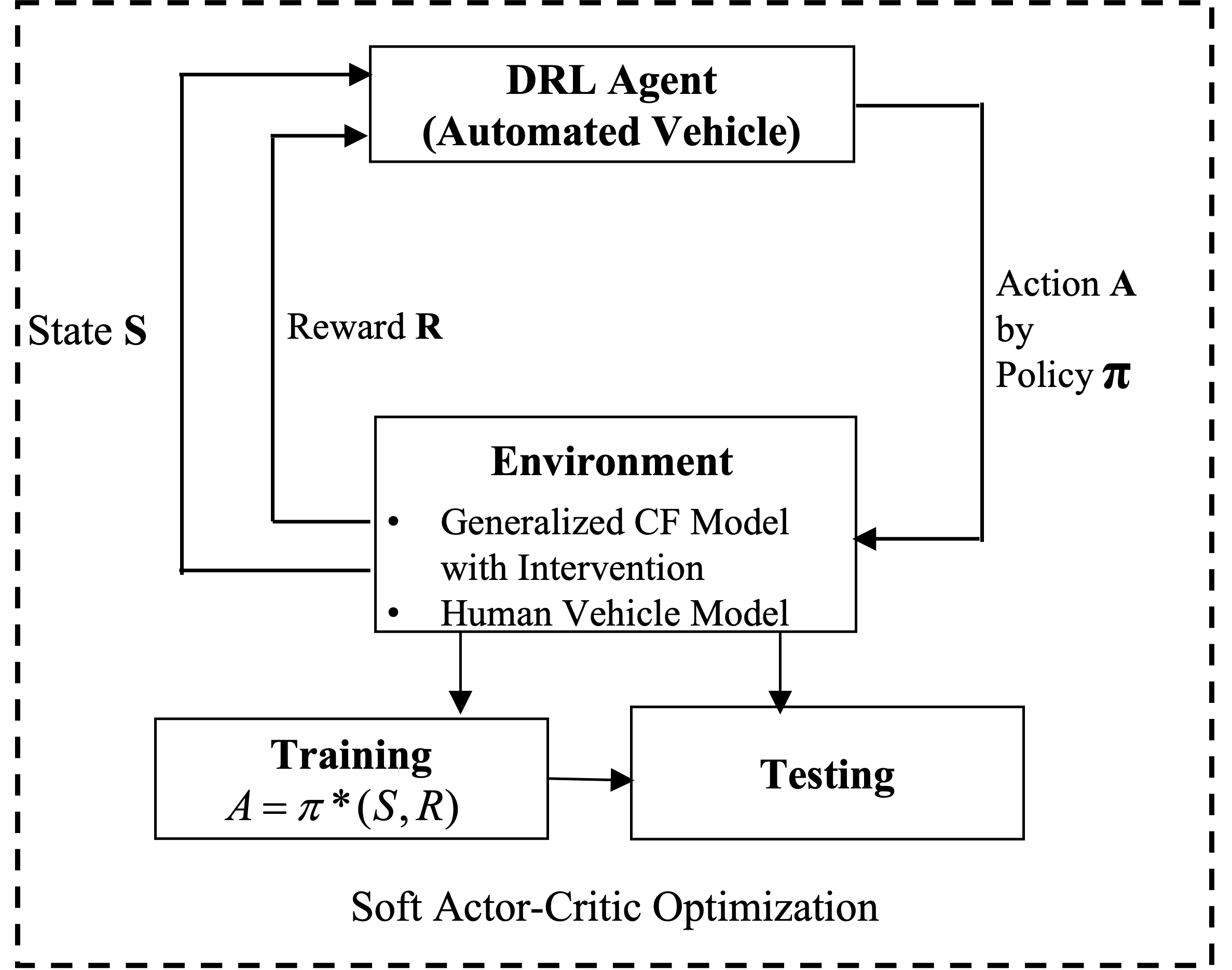}
	\caption{DRL-based Control Framework}
	\label{control_framework}
\end{figure}

The proposed DRL-based controller includes two interactive objects (agent (AV) and environment) and four basic elements (state, policy, reward, and action); see Fig. \ref{control_framework}. The \emph{state} represents the current vehicle kinematics and the driving dissimilarity. The goal is to learn the optimal control \emph{policy} $\boldsymbol{\pi}^*$ that maximizes \emph{reward} that involves  driver intervention and traffic stability. At \emph{state} $\boldsymbol{S}_i(t)$, AV $i$ executes \emph{action} $\boldsymbol{A}_i(t)$ according to \emph{policy} $\boldsymbol{\pi}$, then transitions to a new state $\boldsymbol{S}_i(t+1)$, and receives reward $\boldsymbol{R}_i(t)$. There are three main components in the environment: (1) AV control model (control target), (2) the generalized CF model (Eq. (\ref{CF_law}) with intervention built upon the EA model (Eq. (\ref{EA})), and (3) human driving model. 

We define a three-vehicle platoon as one control unit in the environment, where vehicle $i$ is the AV that is the target of our controller, whose kinematics follow Eq. (\ref{CF_law}), vehicle $i-1$ is the leading HDV from the driving simulator experiment, and vehicle $i+1$ is a HDV whose kinematics can be described by $f_{i+1,HDV}$. Without loss of generality, the kinematics of vehicle $i-1$, $[y_{i-1},\dot{y}_{i-1},\ddot{y}_{i-1}]^T$, represent the downstream traffic condition. Vehicle $i+1$ is included to incorporate the impact of AV control on the upstream traffic. This control unit composition (HDV-AV-HDV) is uniquely designed to incorporate traffic-level insights -- what happens upstream and downstream of the platoon. 

For each control unit, the \emph{state} $\boldsymbol{S}_i(t)$ consists of a vector of spacing, relative speeds of AV $i$ and follower $i+1$ and driving dissimilarity between AV $i$ and its HDV counterpart $\boldsymbol{S}_i(t)=[{y}_{i-1}(t)-{y}_i(t),{y}_{i}(t)-{y}_{i+1}(t), {\dot y}_{i-1}(t)-{\dot{y}}_i(t), {\dot y}_{i-1}(t)-{\dot{y}}_i(t),{w_1}_i\Phi_{i,x}(t), {w_2}_i\Phi_{i,v}(t), {w_3}_i \Phi_{i,TTA}(t)]^T$. The \emph{action} $\boldsymbol{A}_i(t)$, the output of the AV controller, contains acceleration $\ddot{y}_i(t)$ at each time step. The control policy, $\boldsymbol{\pi}$, is an implicit function mapping the states to a probability distribution over the actions. The optimal policy, $\boldsymbol{\pi}^*$, is discovered based on the user-defined reward, $\boldsymbol{R}_i(t)$, during the training process. Here the reward is formulated based on our desired control objectives to (1) minimize voluntary driver intervention and (2) stabilize traffic flow. For (1), we seek to minimize the driving dissimilarity, $U_i(t)$, from Eq. (\ref{Dissimilarity}), as it is equivalent to minimizing driver intervention. For (2), according to $L_2$ string stability \citep{naus2010string,shi2021connected,ploeg2013lp}, we aim to stabilize traffic flow by minimizing two amplifying ratios, which are measured by the ratios of the $L^2$ norms of speed errors between AV $i$ and its leader $i-1$, $\frac{\|\dot{y}_i(t) - \dot{y}_E\|_2}{\|\dot{y}_{i-1}(t) - \dot{y}_E\|_2}$, and between HDV $i+1$ and AV $i$, $\frac{\|\dot{y}_{i+1}(t) - \dot{y}_E\|_2}{\|\dot{y}_{i}(t) - \dot{y}_E\|_2}$, respectively. These two objectives can be conflicting. We want to control the AV to drive similarly to HDVs to minimize driver intervention. Yet, this could come at the cost of unstable behavior by the AV (i.e., disturbance amplification), as HDVs are notoriously string unstable. Besides, we enforce compliance with road regulations, including safety measures and speed limits, with stringent penalties. Thus, we seek to strike a balance. Accordingly, the immediate reward $\boldsymbol{R}_i(t)$ is formulated as: 

\begin{flalign}
    &\boldsymbol{R}_i(t)= -\Bigl[{w_R}_1U_i(t)+{w_R}_2\varrho_1\frac{\|\dot{y}_i(t) - \dot{y}_E\|_2}{\|\dot{y}_{i-1}(t) - \dot{y}_E\|_2}+{w_R}_2\varrho_2\frac{\|\dot{y}_{i+1}(t) - \dot{y}_E\|_2}{\|\dot{y}_{i}(t) - \dot{y}_E\|_2} + \varrho_3(\Delta x_i(t)) +\varrho_4(\dot{y}_i(t))\Bigl] 
\end{flalign}
where ${w_R}_1$, ${w_R}_2$, and ${w_R}_3$ are weight coefficients of the reward function. $\varrho_1$ and $\varrho_2$ are the penalties for the instabilities. $\varrho_3$ is the hard safety penalty for the non-positive spacing $\Delta x_i(t)$, $\varrho_4$ is the penalty for $\dot{y}_i(t)$ exceeding the speed limit.
\begin{flalign}
 \varrho_3(\Delta x_i(t)) =
\begin{cases} 
    \varrho_3& \Delta x_i(t) \leq 0 \\
    0 & \Delta x_i(t) > 0 
\end{cases}
\end{flalign}

\begin{flalign}
 \varrho_4(\Delta x_i(t)) =
\begin{cases} 
    \varrho_4& \dot{y}_i(t) \geq 60 mph \\
    0 &  \dot{y}_i(t) < 60 mph 
\end{cases}
\end{flalign}

The multi-objective control problem is converted to a $\zeta$-discounted future reward problem. Then the optimal control policy ${\boldsymbol{\pi}^*}$ is learned by maximizing the cumulative reward function and the expected entropy of the policy over the state marginal of the trajectory distribution, $\rho_{\boldsymbol{\pi}}(\boldsymbol{S}_i(t))$ as in Eq. (\ref{max_reward}).

\begin{flalign}\label{max_reward}
&\boldsymbol{\pi}^*=\arg\max_{\boldsymbol{\pi}}\sum_{t}\mathbb{E}_{(\boldsymbol{S}_i(t),\boldsymbol{A}_i(t)) \sim \rho_{\boldsymbol{\pi}}}\Bigl\{\zeta^{t}\Bigl[\boldsymbol{R}_i(\boldsymbol{S}_i(t),\boldsymbol{A}_i(t))+\upsilon \mathbb{E}[-log(\boldsymbol{\pi}(\cdot|\boldsymbol{S}_i(t)))] \Bigl]\Bigl\}
\end{flalign}
where $\zeta \in [0,1]$, $\upsilon$ is an automatically tuned temperature parameter that controls the stochasticity of the optimal policy $\boldsymbol{\pi}^*$.

We applied the soft actor-critic (SAC) \citep{haarnoja2018soft} method to accelerate the multi-objective optimization problem solving.  SAC is an actor-critic algorithm, where each action executed by the actor (AV agent) will be appraised by the critic. The policy will be updated in the direction suggested by the critic. Specifically, the critic in SAC suggests that the control policy for AV, $\boldsymbol{\pi}$, should be updated by minimizing the function $J_{\boldsymbol{\pi}}(\phi)$  in Eq. (\ref{new_policy}) at each iteration. The capability of the stable convergence within a highly stochastic environment makes it outperform most other state-of-the-art optimization algorithms (e.g., proximal policy optimization (PPO) or distributed PPO \citep{heess2017emergence,shi2021connected}, deep deterministic policy gradient (DDPG) \citep{lillicrap2015continuous}, etc.). Readers could refer to \cite{haarnoja2018soft} for details about the SAC optimization structure.

\begin{flalign}\label{new_policy}
&J_{\boldsymbol{\pi}}(\phi)= \mathbb{E}[log (\boldsymbol{\pi}_{\phi}(\boldsymbol{A}_t|\boldsymbol{S}_t))-Q_\iota(\boldsymbol{A}_t, \boldsymbol{S}_t)]
\end{flalign}\label{optimization}
where ${\phi}$ denotes the policy parameter, $Q_\iota$ represents a Q-function employed to evaluate the state-action pair, and $\iota$ signifies its associated parameter.

Next, we describe the implementation of the DRL-based control framework, beginning with the identification of key models in the environment. We extend the generalized CF law in Eq. (\ref{CF_law}) into a stochastic form. This extension serves two primary purposes: it captures the complex characteristics intrinsic to real-world human drivers and also augments the training process of the DRL-based controller, providing it with an increased influx of data. Specifically, the following stochastic elements are considered: (1) $D(E(0), d, E^{T}, {w_1},{w_2},{w_3})$: the joint probability distribution of the parameters of the intervention model (i.e., EA model); and (2) $D_{HDV}(\theta)$: the joint probability distribution of the parameters for HDV CF model. By including (1), $E_i(0)$, $d_i$ , $E_i^{T}, {w_1},{w_2},{w_3}$ in Eq. (\ref{EA}) should be drawn from $D(E(0), d, E^{T}, {w_1}, {w_2},{w_3})$. Then the generalized CF model in the environment is determined by Eq. (\ref{refined_CF_AV}).

\begin{flalign}\label{refined_CF_AV}
    \ddot{y}_i(t)&=(1-\eta_i(t))f_{i,HDV}(y_{i-1}(t),\dot{y}_{i-1}(t),y_{i}(t),\dot{y}_{i}(t),D_{HDV}(\theta))+\eta_i(t)f_{i,AV}(t)
\end{flalign}
where, $f_{i,AV}(t)$ is regulated by $\boldsymbol{\pi}^*$, optimized in the DRL-based control. More specifically, we implement, $D(E(0), d, E^{T}, {w_1},{w_2},{w_3})$, which is the posterior joint distribution calibrated using the data from the driving simulator experiment. $D_{HDV}(\theta)$ is obtained by calibrating the stochastic hybrid CF model \citep{jiang2023generic} using real-world HDV data (NGSIM \citep{NGSIM}). 

\section{Control Performance}\label{performance}
This section provides a validation of the proposed longitudinal control mechanism, tested across a robust sample over 2,000 numerical simulations. To better illustrate its capability in managing multi-objective optimization tasks, we benchmark our control against (1) IDM-PID control (used in the driving simulator) and (2) Higher-order linear (HL) control \citep{zhou2020stabilizing} that is stochastically calibrated using real-world AV data \citep{li2021car} by ABC \cite{beaumont2002approximate}. 
Note that we compare with the HL control \citep{zhou2020stabilizing} to validate the performance in a broader and more realistic context, as the driving simulator experiment is deterministic and regulated by the IDM-PID. The HL controller has shown notable robustness in replicating the behaviors of real-world AVs, outperforming other controllers (e.g., lower-order linear (LL) feedback controller, lower-order linear feedback controller with constant spacing policy (LLCS), and model predictive controller (MPC), etc.) \citep{jiang2023generic}. The generalized CF law for Eq. (\ref{CF_law}) is modified into Eq. (\ref{refined_CF_law}) to include the HL model as a comparable model.

\begin{flalign}\label{refined_CF_law}
\begin{aligned}
    \ddot{y}_i(t)&=(1-\eta_i(t))f_{i,HDV}(y_{i-1}(t),\dot{y}_{i-1}(t),y_{i}(t),\dot{y}_{i}(t),D_{HDV}(\theta))\\
    &+\eta_i(t)f_{i,AV}(y_{i-1}(t),\dot{y}_{i-1}(t),y_{i}(t),\dot{y}_{i}(t),D_{AV}(\theta))
\end{aligned}
\end{flalign}
where, $D_{AV}(\theta)$ is the joint probability distribution of parameters in HL model and others have been defined previously. 

The numerical experimental setting is as follows. (1) We utilize the same leading vehicle trajectory, generated from the driving simulator experiment (Fig. \ref{fig_automated}). (2) Vehicle $i$ (AV) and $i+1$ (following HDV) are assumed to start from an equilibrium state with their respective standstill spacing to maintain the equilibrium speed $\dot{y}_E$. To test our control policy, we use cross-validation scheme based on an 80\%/20\% split between training and testing. Data for the training environment come from three parts: (i) empirical driving data, (ii) accepted particles for the EA model, and (iii) accepted particles for the hybrid CF model. Accordingly, 80\% of the data or calibrated parameters, in each of the three parts, are randomly chosen for the training environment. The remaining 20\% are used for testing environment. 
The key distinction between the training and testing environments lies in the policy update process. Specifically, in the testing environment, the policy remains static and does not undergo any further modifications or updates. This differentiation ensures that the policy learned in the training can be assessed accurately based on its performance during the testing. 

\begin{figure*}[!htb]
        \centering
\captionsetup{justification=centering}
        \begin{subfigure}[b]{0.45\textwidth}
 \includegraphics[width=0.9\textwidth]{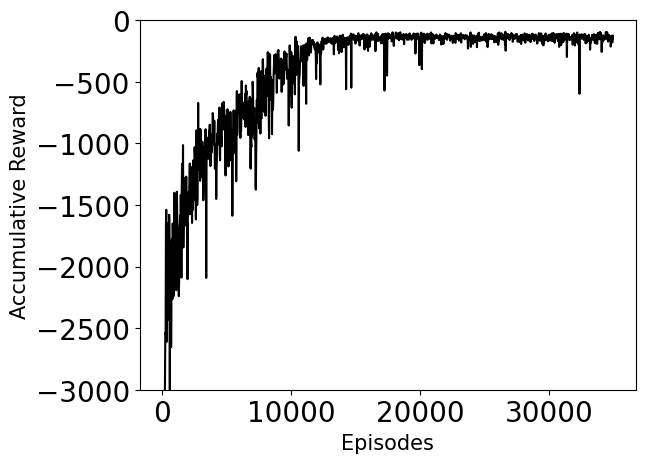}
  \caption*{(a)}%
            {{\small }}   
        \end{subfigure}
        \begin{subfigure}[b]{0.45\textwidth} 
        \centering \includegraphics[width=0.9\textwidth]{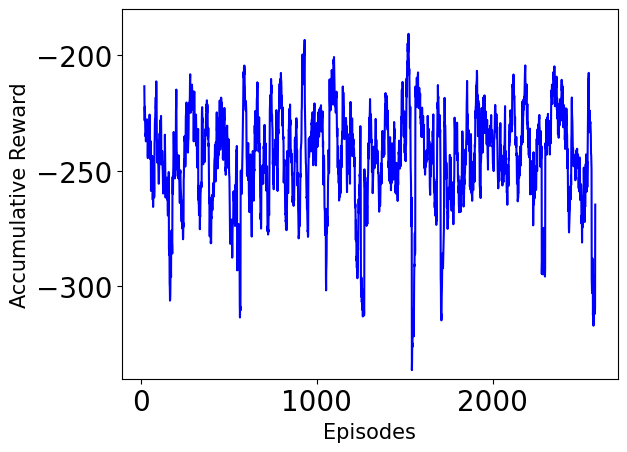}
             \caption*{(b)}%
            {{\small}}   
        \end{subfigure}
         \caption[ Learning Performance]%
        {\small Learning Performance\\
  (a) In Training Environment  (b) In Testing Environment (moving average window: 20)}
        \label{learning}
    \end{figure*}

Table 1 outlines the parameter setting in the training process. Fig. \ref{learning}(a) shows the convergence of the cumulative reward returned in each episode throughout the training process.  Fig. \ref{learning}(b) presents the cumulative rewards attained in the testing environment. Notably, the model reaches convergence after approximately 20,000 episodes, with the moving averaged rewards consistently falling within the range of -340 to -180.  These results suggest that the AV agent has likely learned a near-optimal policy. It is possible that we have not reached a global optimum (or even a local optimum). 
However, we demonstrate below that this policy is still effective in mitigating driver intervention and improving platoon stability.  

\begin{table}[!htp]
\small
	\caption{Parameter Setting in the SAC}\label{tab:versions}
	\begin{center}
		\begin{tabular}{c|c|c|c }\Xhline{1pt}
	    Parameter & Value &Parameter & Value \\\Xhline{1pt}
    Actor learning rate  & $10^{-5}$ & Critic learning rate & $10^{-5}$\\\hline
    $\zeta$&  0.90  & Hidden layer size & 256 \\\hline
    Reward scale & 0.75 & Updated cycles & 25 \\\hline
    Network update interval & 10 & Time step  & 0.1s  \\\hline
    \multicolumn{3}{c|}{Maximum and minimum values for the logarithm of the standard deviation}&2.5, -20\\\hline
    Batch size & 8196 & ${w_R}_1$ & 1 \\\hline
    ${w_R}_2$,${w_R}_3$ & 0.5 &$\varrho_1$ & 1.5 \\\hline
    $\varrho_2$ &1 &$\varrho_3$,$\varrho_4$ & 5  \\\hline
    $D_T$ & 1350 m & $TTT_A$ & 68s\\\Xhline{1pt}	
		\end{tabular}
	\end{center}
\end{table}\label{sac}

\begin{figure*}[!htb]
        \centering
        \captionsetup{justification=centering}
        \begin{subfigure}[b]{0.45\textwidth}
            \centering
   \includegraphics[width=0.9\textwidth]{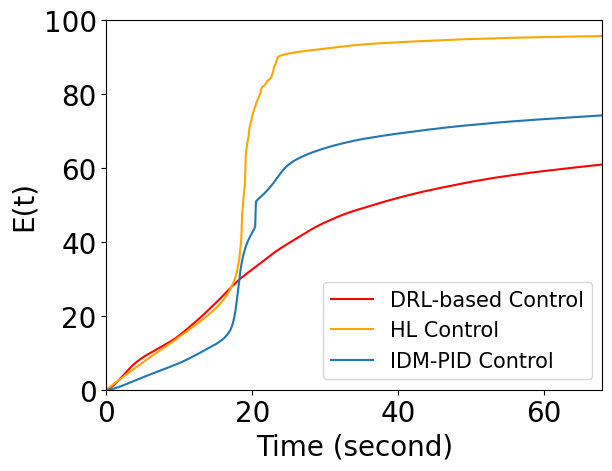}
              \caption[]%
            {{\small }}   
        \end{subfigure}
         \hspace{1em}
        \begin{subfigure}[b]{0.45\textwidth} 
            \centering 
 \includegraphics[width=0.9\textwidth]{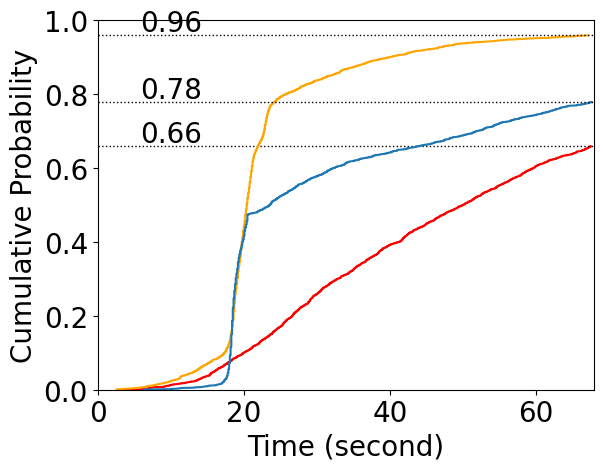}
                 \caption[]%
            {{\small }}   
        \end{subfigure}
         \caption[DRL-Based Control Performance over 2000 Testing Simulations]%
        {\small DRL-Based Control Performance \\
        (a) Expectation of Evidence Accumulation over 2000 Simulations (b) CDF of Takeover Time
        }
        \label{control_performance}
\end{figure*}

\begin{figure*}[!htb]
        \centering
\captionsetup{justification=centering}
        \begin{subfigure}[b]{0.45\textwidth}
 \includegraphics[width=0.9\textwidth]{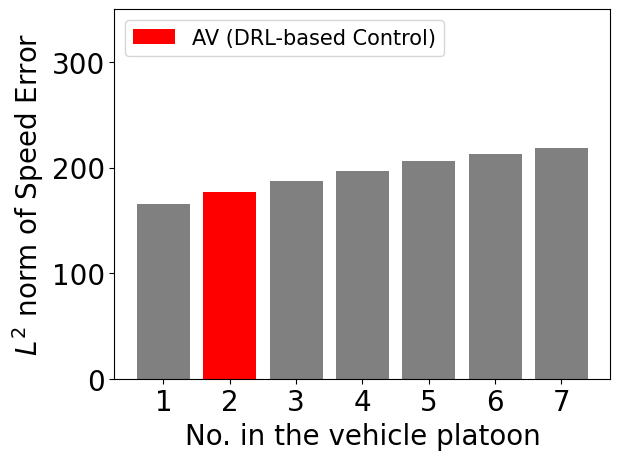}
              \caption[]%
            {{\small }}   
        \end{subfigure}
     \begin{subfigure}[b]{0.45\textwidth} 
            \centering \includegraphics[width=0.9\textwidth]{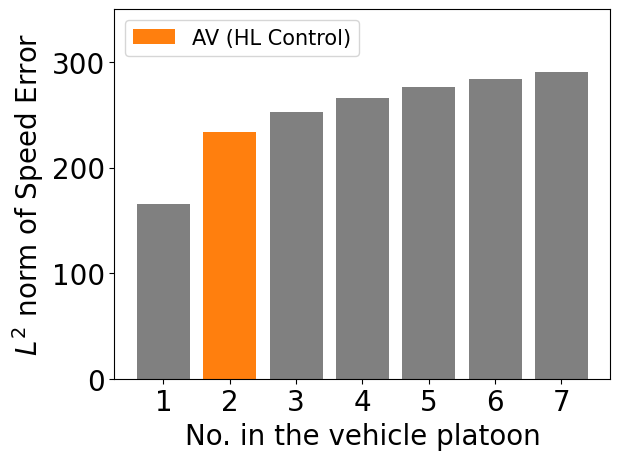}
                 \caption[]%
            {{\small }}   
        \end{subfigure}
        \begin{subfigure}[b]{0.45\textwidth} 
         \centering
            \includegraphics[width=0.9\textwidth]{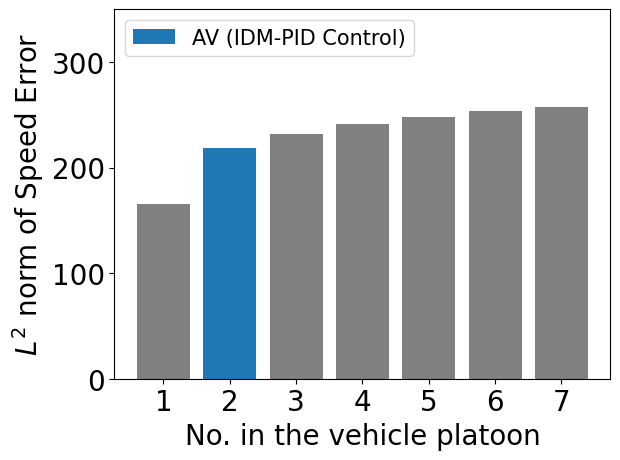}
                 \caption[]%
            {{\small }}   
        \end{subfigure}
                \begin{subfigure}[b]{0.45\textwidth} 
         \centering
            \includegraphics[width=0.9\textwidth]{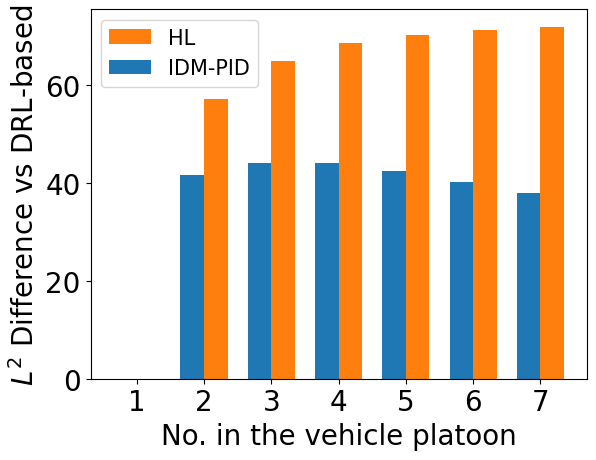}
                 \caption[]%
            {{\small }}   
        \end{subfigure}
         \caption[ $L_2$ norm of Acceleration]%
        {\small Expectation of $L_2$ Norm of Speed Error over 2000 Simulations \\(a) DRL-based Control (b) IDM-PID Control (c) HL Control}
        \label{control_stability}
    \end{figure*}

Figs. \ref{control_performance} and \ref{control_stability} show the overall performance of the DRL-based control over 2000 simulations in the testing environment. In each simulation, we add four HDVs in the platoon, using the stochastic hybrid CF model, to examine disturbance propagation at the platoon-level. 

Fig. \ref{control_performance}(a) shows the expectation of the evidence accumulation across the 2000 simulations. Clearly, the evidence towards distrust accumulates much more slowly under the DRL-based controller than the IDM-PID controller or the HL controller. Fig. \ref{control_performance}(b) shows the cumulative probability function (CDF) of the takeover time. It demonstrates that the DRL-based control results in a notable reduction in intervention events -- over 12\% and 30\% fewer compared to IDM-PID and HL controls, respectively. Both results indicate that the DRL-based control can effectively emulate human-driver CF behavior. In Fig. \ref{control_stability}, we quantify the disturbance magnitude via the expectation of $L^2$ norm of speed error. The disturbance magnitude under the DRL-based controller (Fig. \ref{control_stability}(a)) remains stable at smaller values compared to the IDM-PID (Fig. \ref{control_stability}(b)) and the HL (Fig. \ref{control_stability}(c)) controllers.  Fig. \ref{control_stability}(d) illustrates the notable reduction in disturbance magnitude achieved by the DRL-based control. The results collectively suggest that the proposed DRL-based control effectively handles both objectives to reduce voluntary driver intervention and stabilize the traffic flow at the same time.

\begin{figure*}[!htb]
        \centering
\captionsetup{justification=centering}
            \centering
    \includegraphics[width=\textwidth]{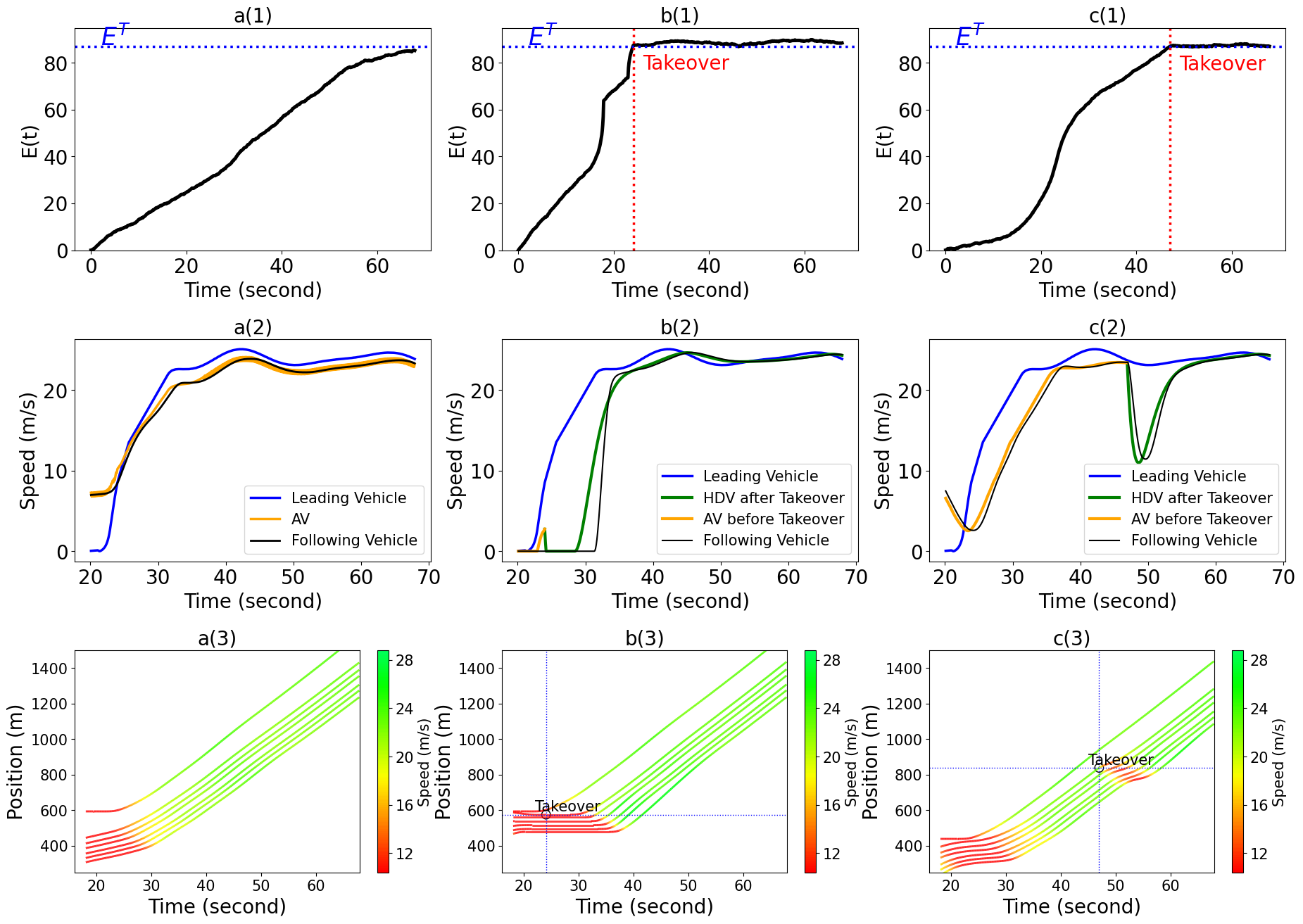}
            \caption*{}%
            {{\small}}    
            \label{}
   \caption[{Impact on Vehicle Kinematics for the Control Unit}]
        {\small Control Performance\\
        (a)  DRL Control without takeover  (b)  IDM-PID Control with Takeover (c)  HL Control with Takeover} 
        \label{control_unit_1}
\end{figure*}

\begin{figure*}[!htb]
        \centering
\captionsetup{justification=centering}
            \centering
    \includegraphics[width=\textwidth]{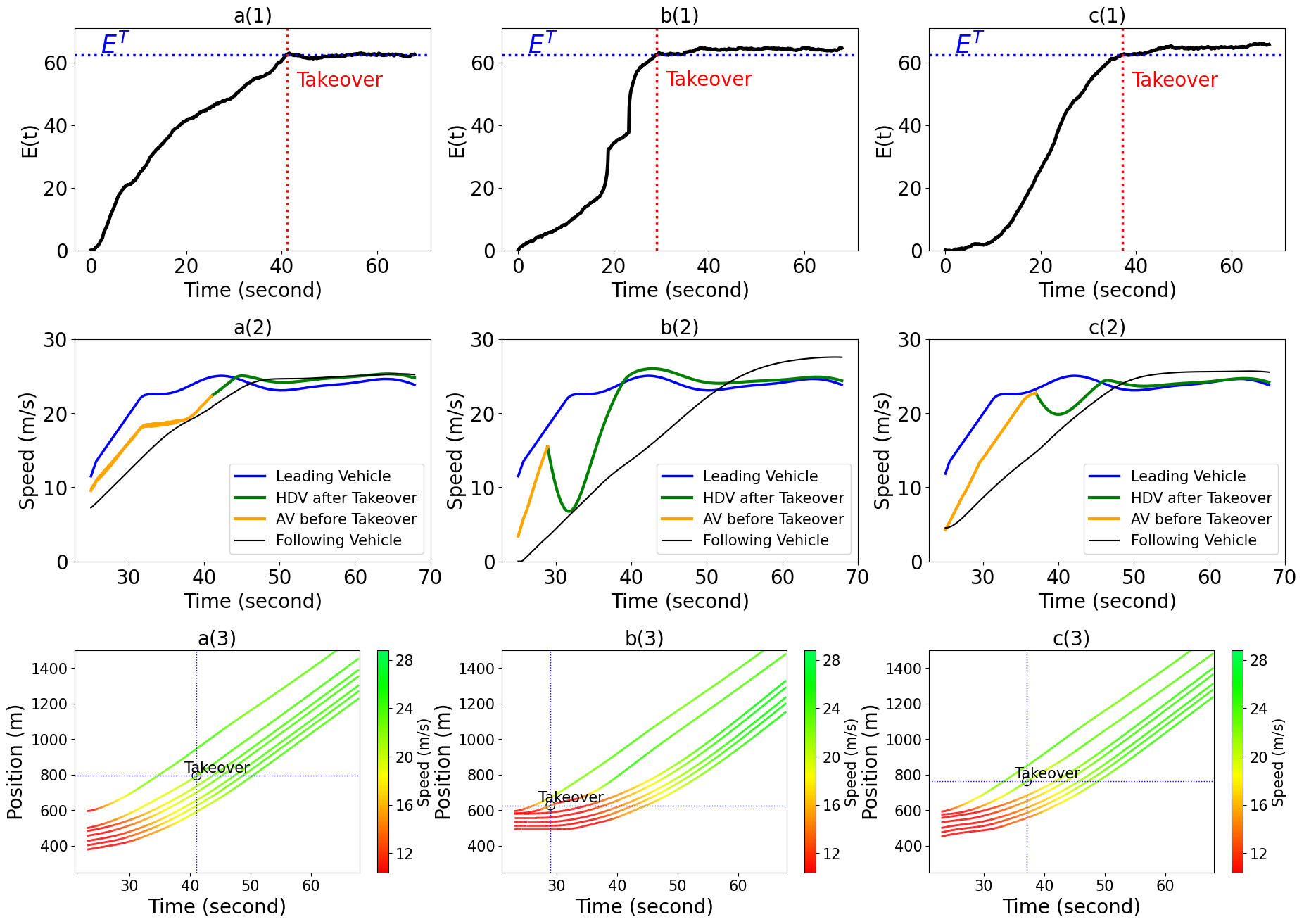}
            \caption*{}%
            {{\small}}    
            \label{}
   \caption[{Impact on Vehicle Kinematics for the Control Unit}]
        {\small Control Performance\\
        (a)  DRL Control with takeover (b)  IDM-PID Control with Takeover (c)  HL Control with Takeover} 
        \label{control_unit_2}
\end{figure*}

Figs. \ref{control_unit_1} and \ref{control_unit_2} show two typical examples of the DRL-based control performance in terms of accumulation of distrust (a(1)-c(1)), vehicle speed (a(2)-c(2)), and disturbance evolution (a(3)-c(3)), as compared to the HL control and the IDM-PID control. As shown in Fig. \ref{control_unit_1}, driver intervention is prevented with the DRL-based control (Fig. \ref{control_unit_1}(a(1))), while it occurs at $24.1s$ with the HL control (Fig. \ref{control_unit_1}(b)) and $47s$ with the IDM-PID control (Fig. \ref{control_unit_1}(c)). Further, the speed is much smoother under the DRL-based control for AV $i$ and the follower $i+1$.  With the (DRL)-based controller, disturbances are mitigated without intervention (Fig. \ref{control_unit_1}(a(3))). In contrast, a noticeable disturbance is observed right after the takeover, which propagates to the follower, with the HL and IDM-PID controllers (Fig. \ref{control_unit_1}(b(3)) and Fig. \ref{control_unit_1}(c(3)), respectively). Fig. \ref{control_unit_2} illustrates an example of interventions occurring across the three controllers (at 41.1s in the DRL-based control in Fig. \ref{control_unit_2}(a), at 29s in the HL control in Fig. \ref{control_unit_2}(b), and at 37.1s in the IDM-PID control in Fig. \ref{control_unit_2}(c)). The transition observed in the DRL-based control is markedly smoother compared to the other two methods in Fig. \ref{control_unit_2}(a(2)-c(2)). This smoothness effectively prevents the amplification of disturbances in the upstream flows in Fig. \ref{control_unit_2}(a(3)-c(3)).

\section{Conclusions}\label{conclusion}

In this paper, we analyze the human-AV interactions through studying voluntary driver intervention, and its implications on traffic performance. Our approach aims to model the causes of voluntary driver intervention, and then design an AV controller that can limit those by controlling its driving behavior. 

We first model the decision-making process of voluntary driver intervention with an evidence accumulation model that describes the evolution of the driver distrust in the AV behavior. Informed through the EA model, and human-in-the-loop driver simulations, we show how in most cases human driver intervention in AVs instigates substantial traffic disturbances that are amplified along the traffic upstream.

In light of this, we propose a DRL-based CF control for AVs that systematically reduces voluntary driver interventions and improves traffic stability. We demonstrate the effectiveness of our model and its superiority against other controllers. Most notably, our model can reduce interventions by a margin of over 12\% $\sim$ 30\% and effectively dampen disturbances. 

A few open questions remain unexplored in this work, and motivate further research in this direction. The current work is limited to driving simulator experiments. Further validation using real-world data is desired. There remain fundamental advances to be made in learning an optimal policy in presence of large action and state spaces, and complex environmental interactions. We hope to motivate further work on the complex nature of human-AV interaction and proper control strategies for traffic enhancement.

\section{Acknowledgements}
This research was sponsored by the US National Science Foundation (Award CNS 1739869).


\bibliographystyle{elsarticle-harv}
\biboptions{authoryear}
\bibliography{references.bib}
\end{document}